%% file: santini_astroph.tex
\documentclass[traditabstract]{aa}
\usepackage{graphicx}
\usepackage{txfonts}
\usepackage{natbib}
\usepackage{enumitem}
\bibpunct{(}{)}{;}{a}{}{,}

 \newcommand{\mic}{$\mu$m}
 \newcommand{\md}{$\rm M_{dust}$}
 \newcommand{\ms}{$\rm M_{star}$}
 \newcommand{\mg}{$\rm M_{gas}$}
 \newcommand{\msbin}{$\log \rm M_{star}[M_\odot]$}
 
 \newcommand{\sfrbin}{$\log \rm SFR[M_\odot/yr]$}
 \newcommand{\fgas}{$\rm f_{gas}$}

\begin{document}
\title{The evolution of the dust and gas  content in galaxies} 

\author{
P.~Santini\inst{1}
\and
R.~Maiolino\inst{2,3}
\and
B.~Magnelli\inst{4}
\and
D.~Lutz\inst{5}
\and
A.~Lamastra\inst{1}
\and
G.~Li Causi\inst{1} 
\and
S.~Eales\inst{6}
\and
P.~Andreani\inst{7,8}
\and
S.~Berta\inst{5}
\and
V.~Buat\inst{9}
\and
A.~Cooray\inst{10}
\and
G.~Cresci\inst{11}
\and
E.~Daddi\inst{12}
\and
D.~Farrah\inst{13}
\and
A.~Fontana\inst{1}
\and
A.~Franceschini\inst{14} 
\and
R.~Genzel\inst{5}
\and 
G.~Granato\inst{8}
\and
A.~Grazian\inst{1}
\and
E.~Le Floc'h\inst{12}
\and
G.~Magdis\inst{15}
\and
M.~Magliocchetti\inst{16}
\and
F.~Mannucci\inst{17}
\and
N.~Menci\inst{1}
\and
R.~Nordon\inst{18}
\and
S.~Oliver\inst{19}
\and
P.~Popesso\inst{5,20}
\and
F.~Pozzi\inst{21}
\and 
L.~Riguccini\inst{22,23}
\and 
G.~Rodighiero\inst{14}
\and
D.~J.~Rosario\inst{5}
\and
M.~Salvato\inst{5}
\and
D.~Scott\inst{24}
\and
L.~Silva\inst{8}
\and 
L.~Tacconi\inst{5}
\and
M.~Viero\inst{25}
 \and
L.~Wang\inst{26}
 \and
S.~Wuyts\inst{5}
 \and
K.~Xu\inst{27}
}

 \offprints{P. Santini, \email{paola.santini@oa-roma.inaf.it}}

\institute{INAF - Osservatorio Astronomico di Roma, via di Frascati 33, 00040 Monte Porzio Catone, Italy
\and Cavendish Laboratory, University of Cambridge, 19 J. J. Thomson
Ave., Cambridge CB3 0HE, UK
\and Kavli Institute for Cosmology, University of Cambridge, Madingley Road, Cambridge CB3 0HA, UK
\and Argelander Institute for Astronomy, Bonn University, Auf dem H\"{u}gel 71, D-53121 Bonn, Germany
\and Max-Planck-Institut f\"{u}r Extraterrestrische Physik (MPE), Postfach 1312, 85741 Garching, Germany
\and School of Physics and Astronomy, Cardiff University, Queens Buildings, The Parade, Cardiff CF24 3AA, UK
\and ESO, Karl-Schwarzschild-Str. 2, D-85748 Garching, Germany
\and INAF - Osservatorio Astronomico di Trieste, via Tiepolo 11, 34131 Trieste, Italy
\and Aix-Marseille Universit\'e, CNRS – LAM (Laboratoire d’Astrophysique de Marseille) UMR 7326, 13388 Marseille, France
\and Department of Physics \& Astronomy, University of California, Irvine, CA 92697, USA
\and INAF - Osservatorio Astronomico di Bologna, via Ranzani 1, 40127 Bologna, Italy
\and Laboratoire AIM, CEA/DSM-CNRS-Universit{\'e} Paris Diderot , IRFU/Service d'Astrophysique, B\^at.709, CEA-Saclay, 91191 Gif-sur-Yvette Cedex, France
\and Department of Physics, Virginia Tech, Blacksburg, VA 24061, USA
\and Dipartimento di Astronomia, Universit\`a di Padova, vicolo Osservatorio, 3, 35122 Padova, Italy
\and Department of Physics, University of Oxford, Keble Road, Oxford OX1 3RH, UK
\and INAF - IAPS, Via Fosso del Cavaliere 100, 00133 Roma, Italy
\and INAF - Osservatorio Astroﬁsico di Arcetri, Largo E. Fermi 5, I-50125 Firenze, Italy
\and School of Physics and Astronomy, The Raymond and Beverly Sackler Faculty of Exact Sciences, Tel Aviv University, Tel Aviv 69978, Israel
\and Astronomy Centre, Department of Physics and Astronomy, University
of Sussex, Brighton, BN1 9QH, UK
\and Excellence Cluster Universe, Boltzmannstr. 2, D-85748 Garching, Germany
\and Dipartimento di Astronomia, Universit{\`a} di Bologna, via
Ranzani 1, 40127 Bologna, Italy
\and NASA Ames REserach Center, Moffett Field, CA 94035, USA
\and BAER Institute, Sonoma, CA, USA
\and Department of Physics \& Astronomy, University of British Columbia, 6224 Agricultural Road, Vancouver, BC V6T 1Z1, Canada
\and California Institute of Technology, 1200 E. California Blvd., Pasadena, CA 91125, USA
\and Department of Physics, Durham University, South Road, Durham, DH1
3LE, UK    
\and NHSC, IPAC, Caltech 100-22, Pasadena, CA 91125, USA
}

   \date{Received .... ; accepted ....}
   \titlerunning{The evolution of the dust and gas content in
     galaxies} 

   \abstract{ We use deep Herschel\thanks{Herschel is an ESA space observatory with science instruments provided
by European-led Principal Investigator consortia and with important
participation from NASA.}  observations taken with both PACS
     and SPIRE imaging cameras to estimate the dust mass of a sample
     of galaxies extracted from the GOODS-S, GOODS-N and the COSMOS
     fields.  We divide the redshift--stellar mass (\ms)--Star
       Formation Rate (SFR) parameter space into small bins and
       investigate average properties over this grid. 
	 In the first part of the work we investigate the
     scaling relations between 
dust mass, 
stellar mass and 
SFR out to $z = 2.5$.
  No clear evolution of the dust mass with redshift is
       observed at a given SFR $and$ stellar mass. 
     We find a tight correlation between the SFR and the dust mass,
     which, under reasonable assumptions, is likely a consequence of
     the Schmidt-Kennicutt (S-K) relation. 
     The previously observed
     correlation between the stellar content and the dust content
     flattens or sometimes disappears when considering galaxies with
     the same SFR. Our finding suggests that most of the correlation between
     dust mass and stellar mass obtained by previous studies is likely
     a consequence of the correlation between the dust mass and the
     SFR combined with the Main Sequence, 
   i.e., the tight relation observed between the stellar mass and
       the SFR and followed by the majority of star-forming galaxies. 
     We then investigate the gas content as inferred from dust mass
     measurements. We convert the dust mass into gas mass by
     assuming that the dust-to-gas ratio scales linearly with the
       gas metallicity (as supported by many observations).
	    For normal star-forming galaxies (on the Main Sequence)
	   the inferred relation between the
     SFR and the gas mass (integrated S-K relation)  broadly agrees 
     with the results of previous studies based on CO measurements,
     despite the completely different approaches.  We observe that all
     galaxies in the sample follow, within uncertainties, the same S-K
     relation.  However, when investigated in redshift intervals, the
     S-K relation shows a moderate, but significant redshift
     evolution.  The bulk of the galaxy population at $z\sim 2$
     converts gas into stars with an efficiency (star formation
     efficiency, SFE=SFR/\mg, equal to the inverse of the depletion
     time) about 5 times
     higher than at $z\sim0$. However, it is not clear what fraction of such
	 variation of the SFE is due to an  intrinsic redshift evolution
	 and what fraction is simply a consequence of
     high-$z$ galaxies having, on average, higher SFR, combined with
     the super-linear slope of the S-K relation  (while other
       studies find a linear slope).
	 We confirm
     that the gas fraction  ($\rm f_{gas} =
       M_{gas}/(M_{gas}+M_{star})$) decreases with stellar mass
     and
 increases with the SFR.  We observe no evolution with
       redshift once \ms~$and$ SFR are fixed. We explain these trends
       by introducing a universal 
     relation between gas fraction, stellar mass and
 SFR  that
     does not evolve with redshift, at least out to $z\sim
     2.5$. 
     Galaxies move across this relation
 as their gas content evolves
     across the cosmic epochs. 
     We use the 3D {\it fundamental $f_{gas}$--$M_{star}$--SFR
       relation}, along with
 the evolution of the Main Sequence with
     redshift, to estimate the evolution
 of the gas  fraction in
     the average population of galaxies as a function of redshift and
     as a function of stellar mass: we find that \ms$\gtrsim 10
       ^{11} \rm
       M_{\odot}$ galaxies show the strongest evolution at $z\gtrsim1.3$
       and a flatter trend at lower redshift, while \fgas~decreases
       more regularly over the entire redshift range probed in \ms$\lesssim 10
       ^{11} \rm M_{\odot}$ galaxies, in agreement with a downsizing
       scenario.
}


\maketitle


\section{Introduction}\label{sec:intro}

Dust is an important component for understanding the galaxy formation
and evolution paradigm. Dust abundance is directly connected with
galaxy growth through the formation of new stars.  Indeed, dust is
composed of metals 
produced
by stellar nucleosynthesis, and then expelled into the interstellar
medium (ISM) via stellar winds and supernovae explosions. A fraction
of these metals mixes with the gas phase, 
while about 30--50\% \citep{draine07} of them condenses into dust
grains. Therefore, dust represents a consistent fraction of the total
mass of metals 
and can be considered as a proxy for the gas metallicity.  While dust
is produced by the past star formation history, it also affects
subsequent star formation, since it enhances the formation
of molecules,  hence allowing the formation of molecular clouds from
which stars are produced. 
Moreover, dust may affect the shape of
the  Initial Mass Function (IMF), through favouring the formation of low-mass stars by
fostering cloud fragmentation in low-metallicity environments and
inhibiting the formation of massive stars \citep{omukai05}.  Finally
dust also affects the detectability of galaxies, because it absorbs
the UV starlight and reradiates it at longer wavelengths.  For all
these reasons, investigating dust properties and dust evolution is a
powerful diagnostic to achieve a more complete view of galaxy
evolution throughout cosmic time.

With the launch of ESA's Herschel Space Observatory
\citep{pilbratt10}, thanks to its improved sensitivity and angular
resolution with respect to previous instruments, it has become
possible to investigate 
dust properties in large samples of galaxies  \citep[e.g.,][and many others]{dunne11,buat12,magdis12,magnelli13,symeonidis13}. Its two imaging
instruments, PACS \citep{poglitsch10} and SPIRE \citep{griffin10},
accurately sample the far-infrared (FIR) and submillimetre dust peak
from 70 to 500~\mic. In this work we use the data collected by two
extragalactic surveys, PEP (PACS Evolutionary Probe, \citealt{lutz11})
and HerMES (Herschel Multi-tiered Extra-galactic Survey,
\citealt{oliver12}), to investigate the evolution of the dust and gas
content in galaxies from the local Universe out to $z\sim 2.5$.

We first study how the
dust content scales with the galaxy stellar content and Star
Formation Rate (SFR). Dust mass, stellar mass and SFR are essential
parameters for understanding the evolution of galaxies. Since dust is
formed in  the atmosphere of evolved stars and in SN winds,
we expect  these parameters  to be 
tightly linked with each other. 
The scaling relations  between dust mass, stellar mass and SFR in the local
or relatively nearby ($z<0.35$) Universe have been investigated by recent
studies based on Herschel data, such as \cite{cortese12} and
\cite{bourne12}. 
In this work, we extend the analysis to higher redshifts, and by
enlarging the Herschel detected sample by means of a stacking
analysis we gain enough statistics to study the correlations between
the dust mass and either the stellar mass or the SFR, by keeping the
other parameter fixed within reasonably small intervals.  For the
first time 
we investigate the dust scaling relations by disentangling the effects
of stellar mass and those of the SFR. This resolves degeneracies
associated with the so-called star formation Main
 Sequence (MS
hereafter). 
 The latter is a
 tight correlation observed between
the SFR and the stellar mass 
 from the local Universe out to at
least $z\sim 3$, with a roughly 0.3 dex scatter \citep[e.g.,][and
references
therein]{brinchmann04,noeske07,elbaz07,santini09,karim11,rodighiero11,whitaker12}.
Galaxies on the MS are thought to form stars through secular processes
by gas
 accretion from the  Integalactic Medium. 
Outliers above the
 MS are defined as starbursts
\citep[e.g.,][]{rodighiero11}. Star formation episodes in these
galaxies are violent and rapid, likely driven by mergers
\citep[e.g.,][]{elbaz11,wuyts11b,nordon12}. Despite the much more vigorous 
star formation activity observed in starbursts, according to recent
studies \citep[e.g.,][]{rodighiero11,sargent12,lamastra13}, these galaxies
play a minor role in the global star formation history of the
Universe, accounting for only $\sim 10\%$ of  the cosmic SFR density at $z\sim 2$. Since at
any redshift most of the galaxies are located
 on the MS, most
studies cannot investigate the dependence of physical quantities
(e.g., dust content) on stellar mass and SFR independently, since these
two
 quantities are degenerate along the MS. To disentangle the
intrinsic
 dependence
 on each of these quantities large samples of
objects are required to properly
 investigate the dependence on SFR
at any fixed \ms~and, viceversa,
 the dependence on \ms~at a
fixed SFR. 

Knowledge of the dust content can be
further exploited to obtain information
 on the gas content, if the
dust-to-gas ratio is known. In the past, most studies
 on the gas
content in high-$z$ galaxies have been
 based on CO observations
\citep[e.g.][]{tacconi10,tacconi13,daddi10,genzel10}.
 These studies
have allowed the investigation of 
 the relation between the
molecular gas mass and the SFR, i.e., the Schmidt-Kennicutt relation
(\citealt{schmidt59,kennicutt98b}, S-K hereafter), 
 at different
cosmic epochs.
 However, these observations are time consuming
  and   affected by 
uncertainties associated with the CO-to-H$_2$ conversion factor,
 which is poorly constrained for starburst or metal-poor galaxies
\citep[see][for a review]{bolatto13}. 

An alternative method to derive the gas content is to exploit the
dust masses inferred from FIR-submm measurements and convert them
into gas masses by assuming a dust-to-gas ratio
\citep[e.g.,][]{eales10,leroy11,magdis11b,scoville12}.
 We adopt this approach in the
second part of this work.  We convert
 the dust mass into gas mass by
assuming that the dust-to-gas ratio scales  linearly with the gas metallicity
and that dust properties are similar to those in the local Universe,
where the method is calibrated.  We estimate
 the gas metallicity
from our data by exploiting the Fundamental
 Metallicity Relation
(FMR hereafter) fitted by \cite{mannucci10} on
 local galaxies and
shown to hold out to $z\sim 2.5$. According to the
 FMR, the gas
metallicity only depends on the SFR and the stellar mass,
 and does
not evolve with redshift (see also \citealt{laralopez10}). 
 With
these assumptions, which will be discussed in the text, we 
study the relation between the SFR and the gas mass and investigate
the 
 evolution of the gas fraction out to $z\sim 2.5$ independently
of CO measurements. 
  We note, however, that the two methods for measuring the gas
  mass (the ``dust-method'' and CO observations) are cross-calibrated
  with each other.

A similar approach was adopted by \cite{magdis12} by using Herschel
data
 from the GOODS-Herschel survey. We improve over their work by
also using the data
 in the COSMOS field that, thanks to the large
number of objects, allows us
 to greatly expand the stacking
technique to a range of galaxy physical parameters
 not explored by
\cite{magdis12}, and to significantly shrink the
uncertainties. Moreover, while \cite{magdis12} bin the data in terms
of their
 distance from the MS at any redshift, we bin our data in
stellar mass, SFR and
 redshift, to avoid the inclusion of any
a-priori relation between stellar mass and SFR and to study the
existing trend as a function of physical parameters.

The paper is organized as follows. After presenting the data set
(Sect. \ref{sec:data}) and the method used to compute SFR, stellar,
dust and gas masses, and gas metallicities (Sect. \ref{sec:method}),
we present the dust scaling relations in Sect. \ref{sec:scaling}, and
the study of the evolution of the gas content in
Sect. \ref{sec:gascontent}.  Finally, we summarize our
 results in
Sect. \ref{sec:summ}.

In the following, we adopt the
$\Lambda$-CDM concordance cosmological
 model (H$_0$ = 70 km/s/Mpc,
$\Omega_M = 0.3$ and $\Omega_{\Lambda} =
 0.7$) and a Salpeter
IMF. 

\section{The data set} \label{sec:data}

For this work we take advantage of the wide photometric coverage
available in three extragalactic fields: the two deep GOODS fields
(GOODS-S and GOODS-N,  $\sim 17'\times11'$ each) and the much
larger but shallower COSMOS 
field ($\sim 85'\times 85'$). Dealing with these fields together represents an excellent
combination of 
having good statistics on both bright and faint sources
from low to high redshift.

Most important for the aim of this work, i.e., essential to derive
dust
 masses, are the FIR observations carried out by Herschel with
the shorter
 wavelength (70, 100 and 160~\mic) PACS camera and the
longer
 wavelength (250, 350, 500~\mic) SPIRE camera. As anticipated
in
 Sect. \ref{sec:intro}, we use the data collected by the two
extragalactic
 surveys PEP and HerMES. 
 Catalogue extraction on
Herschel maps is based on a PSF fitting
 analysis that makes use of
prior knowledge of MIPS 24~\mic~positions
 and fluxes. PACS
catalogues are described in \cite{lutz11} (and references therein)
and \cite{berta11}, while SPIRE catalogues are
 presented in
\cite{roseboom10} and are updated following
 \cite{roseboom12}.  The
$3\sigma$ limits\footnote{In deep 160, 250,
 350 and 500~\mic~
  observations, rms values include confusion noise.} at 100, 160,
250, 350 and 500~\mic~are 1.2, 2.4, 7.8, 9.5, 12.1 mJy
 in GOODS-S,
3.0, 5.7, 9.2, 12.0, 12.1 mJy in GOODS-N and 5.0, 10.2,
 8.1, 10.7,
15.4 mJy in COSMOS, respectively.
 The only field which was observed at 70~\mic~is
GOODS-S. After testing that the use of 70~\mic~photometry does not
introduce any significant difference in the dust mass estimates, we
ignored this band for consistency with the other fields. 

In order to infer redshifts and other properties needed for this
study, we complement Herschel observations with public multiwavelength
photometric catalogues. 
 For GOODS-S we use the updated GOODS-MUSIC
catalogue \citep{santini09,grazian06}. 
 For GOODS-N we use the
catalogue compiled by the PEP Team and
 described in \cite{berta10}
and \cite{berta11}, publicly available at \footnote{http://www.mpe.mpg.de/ir/Research/PEP/GOODSN\_multiwave}.  For the COSMOS field we use the multiwavelength
catalogue  presented in \cite{ilbert09} and \cite{mccracken10} and available at
\footnote{http://irsa.ipac.caltech.edu/data/COSMOS/tables/photometry/}. COSMOS
data reduction is described in \cite{capak07}, although the new
catalogue uses better algorithms for source detection and photometry
measurements. 
 This catalogue is supplemented
 with IRAC photometry
from \cite{sanders07} and \cite{ilbert09} and 24~\mic~
 photometry
from \cite{lefloch09}.

All the catalogues are supplemented with either spectroscopic
 or  photometric redshifts.  Spectroscopic redshifts are available for  $\sim 30\%$, $\sim 27\%$ and
$\sim 3\%$ of the final sample, respectively, in GOODS-S, GOODS-N and
COSMOS.  
 For the remaining sources, we adopt the photometric redshift
  estimates publicly released with the two GOODS catalogues and
those computed by the authors for COSMOS and presented in
\cite{berta11}. The latter were computed for all sources rather than
for the $I$-selected subsample released by \cite{ilbert09}, and show
similar quality for the objects in common. 
Photometric redshifts in GOODS-S are estimated by fitting the multiwavelength photometry
to the PEGASE 2.0 templates \citep{fioc97}, as presented in
\citealt{grazian06} and updated as in \citealt{santini09}. For GOODS-N
and COSMOS, the EAZY code \citep{brammer08} was adopted, as discussed
in \cite{berta11}.  We refer to the papers cited above, as well as to
\cite{santini12b} for more detailed information about spectroscopic
and photometric redshifts and their accuracy.  

\subsection{Sample selection}

In order to achieve a reliable estimate of the main physical
  parameters required for this analysis, we need to apply some
  selections to the galaxy sample in the three fields.

 We firstly  require the signal-to-noise ratio in
$K$ band to be larger than 10. This selection ensures clean photometry
and reliable stellar mass estimates for all sources.  

Secondly, in order to estimate the SFR from an IR tracer,
  independent of uncertain corrections for dust extinction, we require
  a 24~\mic~detection for all galaxies (see Sect. \ref{sec:sfr}). This
  is the tightest  selection criterion and limits the final sample to
  galaxies with relatively high star formation
 (32--52\% of the sample, depending on the field).
However, although it reduces the dynamical range probed, 
a SFR cut is not an issue for most of this study, since we
analyse trends as a function of SFR or at fixed SFR. In the latter case, the
use of narrow SFR intervals prevents strong incompleteness effects within each
individual bin.

Finally, we 
remove all known AGNs from the catalogues ($\sim2.5\%$ of the total
final sample), by considering X-ray detected sources (the AGN
sample of \citealt{santini12b}), highly obscured AGNs detected through
their mid-IR excess \citep[following][]{fiore08}, and IRAC selected
AGNs \citep{donley12}. Indeed, besides the cold dust heated by star
formation regions, these sources host a warm dust component, which is
heated by nuclear accretion processes and which might bias the dust
mass estimates.

\section{Parameters determination} \label{sec:method}

We describe in this section how the basic ingredients of our analysis,
i.e., stellar masses (\ms), SFR, dust masses (\md), gas masses (\mg)
and gas metallicities, are obtained.

\subsection{Stellar masses} \label{sec:mstar}

Stellar masses are estimated by fitting observed near-UV to near-IR
photometry with a library of stellar synthetic templates
\citep[e.g.][]{fontana06}. We adopt the same procedure described in
\cite{santini09}: we perform a $\chi^2$ minimization of \cite{bc03}
synthetic models, parameterizing the star 
formation histories as exponentially declining laws  of timescale
$\tau$ and assuming a Salpeter\footnote{Conversion factors
  to a Chabrier IMF are given in Sect. \ref{sec:gmetmgas}.} IMF. Age, gas
  metallicity,  $\tau$ and reddening are set as free parameters, and
  we use a \cite{calzetti00} or SMC extinction curve (whichever
    provides the best fit). We refer to
  \cite{santini09} and references therein for more details on the
  stellar template library. In the fitting procedure,   
each band is
weighted with the inverse of the photometric uncertainty.  Since
\cite{bc03} models do not include emission from dust reprocessing, we
fit the observed flux densities out to 5.5~\mic~rest-frame. The redshift is
fixed to the photometric or spectroscopic one, where available.

To ensure reliable stellar mass estimates, in the following we remove
all
 sources with a reduced $\chi^2$ larger than 10 ($\sim$ 4--13\% of
the final sample, depending on the field).

Our sample spans a large redshift interval, hence the range of
  rest-frame wavelengths used to measure stellar masses is not the
  same for all sources. More specifically, high-$z$ galaxies lack
  constraints at the longest rest-frame wavelengths. However, \cite{fontana06} have shown
  that the lack of IRAC bands when estimating the stellar mass from
  multi-wavelength fitting, while producing some scatter, does not
  introduce any systematics (see also \citealt{mitchell13}). In any case, the rest-frame $K$ band,
  essential for a reliable stellar mass estimate, is sampled even at
  the highest redshifts probed by our analysis.

\subsection{SFR} \label{sec:sfr}

Star formation rates 
are estimated from the total IR
luminosity
 integrated between 8 and 1000~\mic~($\rm L_{IR}$) and
taking into account
 the contribution from unobscured SF. We use the
calibrations adopted by \cite{santini09} (see references therein):
\begin{eqnarray}
\label{eq:sfr}
\rm SFR[M_\odot/yr]  &  =  &  1.8 \times 10^{-10}\times \rm L_{bol}[L_\odot]; \\
\rm L_{bol} &  = &  2.2 \times \rm L_{UV} + \rm L_{IR}. \nonumber
\end{eqnarray}
Here $ \rm  L_{UV} = 1.5 \times \nu L_\nu({2700\AA})$ is the rest-frame UV
luminosity derived from the SED fitting and uncorrected for
extinction. 

Since Herschel detections
 are only available for $\sim$ 11--25\%
(depending on the field) of the
 sample\footnote{The statistics
    given in this section refers to the sample in the redshift and
    stellar mass range of interest and in the area over which the
    analysis is carried out (see Sect. \ref{sec:stacking}).}, in
order to have a consistent SFR estimate for a larger number of
sources, we estimate $\rm L_{IR}$ from the 24~\mic~MIPS band 
  (reaching 3$\sigma$ flux limits of 20 and 60 $\mu$Jy 
  in the GOODS fields and in COSMOS, respectively).
Most importantly, this approach also avoids any degeneracy with the
dust
 mass estimates, derived from Herschel data.  We fit 24~\mic~
flux densities to the MS IR template derived by \cite{elbaz11} on the basis of
Herschel observations. This template, thanks to an updated treatment
of the MIR-to-FIR emission,  overcomes previous issues related
with the 24~\mic~overestimate of $\rm L_{IR}$ and provides a reliable
estimate of the SFR  for all galaxies (see Fig.~23 of \citealt{elbaz11}). As a further confirmation, in Appendix
  \ref{app:cfrsfr} we compare the 24~\mic-based SFR with that derived
  by fitting the full FIR photometry and find very good
  agreement. This test proves that the adoption of 24~\mic-based SFR
  does not introduce relevant biases in the
  analysis. Most importantly, it provides a SFR estimate that is
  independent of the dust and gas mass measurement and therefore
  allows us to confidently investigate correlations among these quantities. 

\subsection{Stacking procedure} \label{sec:stacking}

Dust masses are computed by means of Herschel observations.  
Only a small fraction of the sources are individually detected by
Herschel, and only less than 10\%\footnote{These fractions refer to
    the sample over which the analysis is performed (see below and
    Sect. \ref{sec:mdust}).} 
fulfill the requirements of good FIR sampling adopted for the dust mass estimate (see
Sect. \ref{sec:mdust}). Therefore, a stacking procedure to estimate the average
flux of a group of sources is needed to perform an
analysis, which is unbiased towards the brightest IR galaxies. 
We describe here how average fluxes for subsamples of
sources are estimated. In the next section we explain how such
subsamples are compiled. 

The stacking procedure adopted in this work is similar to that
described by \cite{santini12b} and also used in
 \cite{rosario12} and
\cite{shao10}. First of all, in each Herschel band we restrict to
the area where the coverage (i.e., integration time) is larger than
half its value at the centre of the image.  This removes the image
boundaries 
where stacking  may be less reliable 
due to the
larger noise level.  For each $z$--\ms--SFR bin containing at least
10 sources and for each Herschel band, we stack\footnote{We use the
  \cite{bethermin10} libraries available at
  http://www.ias.u-psud.fr/irgalaxies/downloads.php.} on the residual
image (i.e., map from which all $3 \sigma$ detected sources have been
subtracted) at the positions of undetected sources (by ``undetected''
we mean below $3\sigma$ confidence level).  Each stamp is weighted
with the inverse of the square of the error map. The photometry on the
stacked PACS images is measured by fitting the PSF, while for SPIRE
images we read the value of the central pixel (SPIRE maps are
calibrated in Jy/beam), which was suggested by \cite{bethermin12} to
be more reliable in the case of clustered sources.  Uncertainties in
the stacked flux densities are computed by means of a bootstrap procedure.
The final average flux density $\overline{S}$ is obtained by combining the
stacked flux ($S_{stacked} $) with the individually detected fluxes
($S_i$) in the same bin:
\begin{equation} 
\overline{S} = \frac{S_{stacked}\times N_{stacked} + \sum_{i=1}^{N_{det}}S_i }{N_{tot}}, 
\label{eq:stacking}
\end{equation} 
 where $N_{stacked}$, $N_{det}$ and $N_{tot}$ are the number of
  undetected, detected and total sources, respectively, in the bin.

  The stacking procedure implicitly assumes that sources in the image
  are not clustered. However, in the realistic case sources 
 can be
  clustered with other sources either included or not included in
  the stacking sample. This effect may result in an overestimation of
  the
 flux in blended sources (see, e.g., \citealt{bethermin12} or
  \citealt{magnelli13}).  Given the lack of information 
 on sources
  below the noise level, it is not straightforward to correct
 for
  this effect. However, if we are able to recognize its occurrence,
  we can ignore the bins where the stacking is affected by confusion.
  For this purpose, an ad hoc simulation has been put into place by
  the
 PEP Team. We briefly recall the 
  basic steps of the 
 simulation, and refer the reader to
  \cite{magnelli13} for a
 more detailed description. Synthetic SPIRE
  fluxes
 were estimated through the MS template of \cite{elbaz11}, 
  given the observed
 redshifts and SFRs, and simulated catalogues and
  maps were produced.
 Whenever we stack on a group of sources on
  real SPIRE maps, we also
 stack at the same positions on the
  simulated maps and obtain a
 simulated average flux density
  ($\overline{S}_{sim}$). We compare
 $\overline{S}_{sim}$ with the
  mean value
 ($\overline{S}_{input}$) of the same flux densities contained
  in the simulated
 catalogue (previously used to create the
  simulated maps).  Following \cite{magnelli13}, if
  $|\overline{S}_{input}-\overline{S}_{sim}|/\overline{S}_{input}>0.5$
  we reject the corresponding bin\footnote{We verified that the trends
    presented in this analysis are independent of the chosen
    threshold.}.  The largest blending effects are seen at low flux densities
  and in the 500 \mic~band, as expected.  The criterion above implies
  rejection of $\sim$ 10\% 
of the stacked fluxes at 250 \mic, $\sim$
  16\% at 350 \mic~and $\sim$ 33\% 
at 500 \mic. 
 We also run our analysis by including these bins, to check  
that their rejection 
does not
  introduce any bias in our results.

\subsection{The $z$--\ms--SFR grid and combination across fields
} \label{sec:grid}

The basis of our stacking analysis is to infer an average dust
  mass for sources showing similar properties. To this aim, we divide
  the redshift--stellar mass--SFR parameter space into small bins, and
  run the stacking procedure on all galaxies belonging to each
  bin. The ranges covered by our grid are 0.05--2.5 in redshift,
  9.75--12 in \msbin~and -0.75--3 in \sfrbin. 
The boundaries of
the bins, listed in Tables \ref{tab:stat1} to \ref{tab:stat5} 
  together with the abundance of sources per bin, are
chosen to provide a fine sampling of the \ms-SFR parameter space and at the same time
to have good statistics in each bin. We adopt bins of 0.25
  dex in \ms~and 0.2 dex in SFR at intermediate \ms~and SFR values, 
  where we have the best statistics, and slightly larger bins at the
  boundaries. This choice strongly
limits the level of incompleteness within each individual bin. 
Incompleteness issues will simply result into bins not populated
  and therefore missing from our grid (e.g.,
  at low \ms~and SFR as redshift increases). 

To combine the different fields, we stack on them 
simultaneously by weighting each stamp
with the relative weight map.  The total number of sources in each bin
and the contribution of each field are reported in Tables
\ref{tab:stat1} to \ref{tab:stat5}. Since the statistics are strongly
dominated by the COSMOS field, we do not expect intrinsic differences
among the fields to significantly affect our results.

For each bin of the grid we  compute the average redshift, \ms~and
SFR of the galaxies belonging to it, and associate these values to the
bin. 
The standard deviations of the distribution of these parameters within
the bin provide the error bars 
associated with the average values.
 

\subsection{Dust masses} \label{sec:mdust}

For a population of dust grains at a given temperature and with a
given emissivity, the dust mass can be inferred from their global
thermal infrared grey-body spectrum and, in particular, by its
normalization and associated temperature. More generally, the dust
thermal emission in galaxies is composed by multiple thermal
components. In order to account for this, we use, as a description of
the dust emission, the spectral energy distribution (SED) templates of
\cite{draineli07}.  In doing so, we implicitly assume that the dust
properties and emissivities of our sources are similar to those of
local galaxies, on which the templates were tested \citep{draine07}.
Such assumption is supported by the lack of evolution in the
extinction curves, at least out to $z\sim4$ \citep{gallerani10}. It is
also supported by the gas metallicity range probed by our sample
($\geq 8.58$, see Sect. \ref{sec:gmetmgas}) and by the recent results
of \cite{remyruyer13}, claiming that the gas metallicity does not have
strong effects on the dust emissivity index. Moreover, our sample is
mostly made of MS galaxies. The \cite{draineli07} model is also based
on the assumption that dust is optically thin, plausibly applicable to
our sample, which does not include very extreme sources such as local
ULIRGs or high-$z$ sources forming a few thousands of solar masses per
year. However, as a sanity check, we also have used the GRASIL model
\citep{silva98} which includes extreme optically thick young starburst
components, and the final results are unaffected (see below).  
  Finally, \cite{galliano11}, by studying the Large Magellanic Cloud,
  found that dust masses may be systematically understimated by
  $\simeq$~50\% when computed from unresolved fluxes. The authors
  ascribe this effect to possible vealing of the cold dust component
  by the emission of the warmer regions. However, this effect would
  only introduce an offset without modifying the main results of this
  analysis.  

\begin{figure}[!t]
\resizebox{\hsize}{!}{\includegraphics[angle=0]{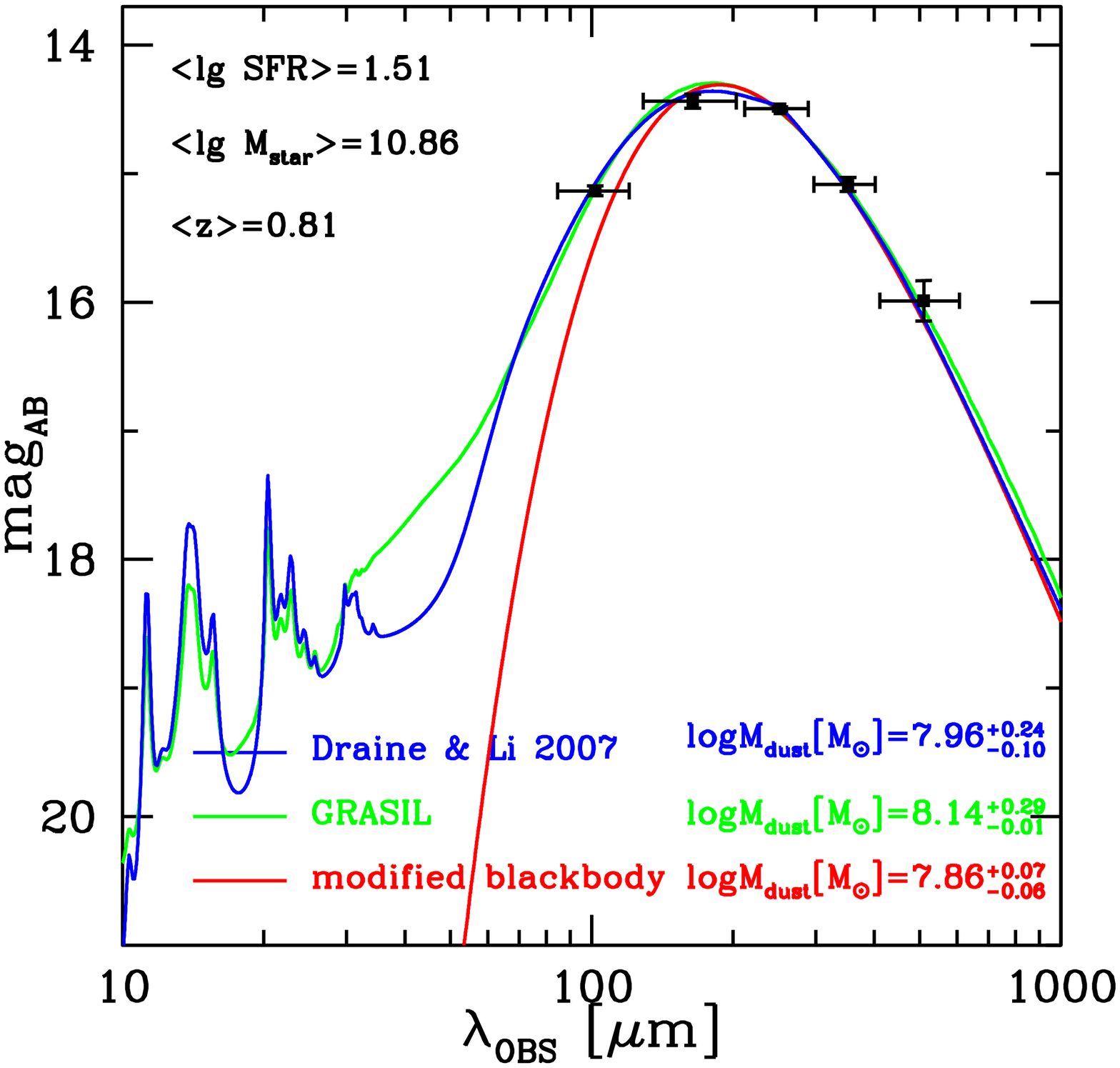}}
\caption{Example of the fits done to estimate the dust mass. Black
  symbols show stacked fluxes in the bin of the $z$--\ms--SFR grid
  with $z=[0.6,1)$, \msbin~$=[10.75,11)$ and \sfrbin~$=[1.4,1.6)$
  The blue line shows the best-fit template from the
 library of
  \cite{draineli07}. For
  a comparison, the green and
 red curves show the fits with the
  GRASIL model and with a
 single-temperature modified blackbody (the latter not fitted to the
   shortest wavelength flux density),
  respectively. The dust mass inferred with  the three libraries is indicated in the bottom
  right corner.  The three libraries differ in the resulting dust
    masses by a roughly constant offset, but yield the same trends. }
\label{fig:bestfit}
\end{figure}

According to
the \cite{draineli07}  model, the interstellar dust is represented as a mixture of
amorphous
 silicate and graphite grains, with size distribution
modeled by
 \cite{weingartner01} and updated as in \cite{draineli07},
mimicking
 different extinction curves. A fraction $q_{PAH}$ of the
total dust
 mass is contributed by PAH particles (with $<1000$ C
atoms).  Although they only provide a minor contribution to the total dust
  mass, their abundance has an important effect in shaping the galaxy SED at
  short wavelengths. The
 majority (a fraction equal to $1-\gamma$) of dust
grains are located
 in the diffuse ISM and heated by a diffuse
radiation field contributed
 by many stars. This results in a single
radiation intensity
 $U=U_{min}$, where $U$ is a dimensionless factor
normalized to the
 local ISM. The rest of the grains are localized in
photodissociation
 regions close to bright stars, and exposed to
multiple and more
 intense starlight intensities
($U_{min}<U<U_{max}$) distributed as a
 power law ($\propto
U^{-\alpha}$).

Following the prescriptions of \cite{draine07}, we build a library of
MW-like models with PAH abundances $q_{PAH}$ in the range 0.47--4.58\%, $0.0<\gamma<0.3$, $\alpha=2$, $U_{max} = 10^6$ and $U_{min}$
varying between 0.7 and 25. This latter prescription (instead of using
$U_{min}\geq 0.1$) prevents the
 risk of fitting erroneous large dust
masses in the absence of rest-frame submillimetre data to constrain the amount
of cool dust.  

Dust masses are derived by fitting and normalizing
the stacked 100-to-500 \mic~Herschel photometry to this template
library.  
The redshift is fixed to
the mean redshift in the bin. The template showing the minimum $\chi^2$ is chosen, and the
normalization of the fit provides a measure of the dust mass.

In the fitting procedure, we require the stacked fluxes to have at
least $3\sigma$ significance. In order to have a good sampling of the
spectrum, especially on the Rayleigh-Jeans side, most sensitive to the dust
mass, we only consider bins with  significant flux in at least 3
  bands, 
of which at least one is longward of rest-frame 160 \mic~\citep{draine07}. This enables to
account for changes in the dust temperature and makes us confident of
the resulting \md. 
26\%
of the total number of bins are rejected because of these selections. 
We visually inspect every single bin to ensure the quality of the
fits,  and conservatively reject 5 of them ($\sim 4\%$), where the stacked fluxes were not
  satisfactorily reproduced by the best-fit template. An example of our fitting output can be seen in
Fig.~\ref{fig:bestfit}. The best-fits for all bins in the final sample
  can be seen in Appendix~\ref{app:mdustfits}.

MIR fluxes are not included in the fit so that the dust mass and SFR
estimates are totally independent.  As a consistency check, we
  also computed \md~by including 24~\mic~flux densities. The resulting
  dust masses are in very good agreement with our reference
  estimates: their mean (median) ratio ($\rm \log \left (
    M_{dust}^{24\mu m}/M_{dust}^{no 24\mu m} \right )$) is
  -0.001 (0.008), with a scatter of 0.07, 
and the average error bar (see below) is only 
  $\sim$10\% lower than without including the 24~\mic~band. This ratio
  shows no trends with either stellar mass, SFR or redshift, except a
  slightly larger scatter at low-$z$ (here rest-frame wavelengths
  below 100~\mic~ are not sampled in the absence of 24~\mic~data).

Errors on \md~are estimated by allowing the stacked photometry to vary
within its uncertainty and the redshift to move around the mean value
within its standard deviation in the bin. The uncertainty is given by the minimum
and maximum \md~allowed by templates whose probability according to a
$\chi^2$ test is larger than 32\%.
All data points whose associated error on \md~is larger than 1 dex
(further $\sim 5 \%$ of the available bins)
are ignored throughout the analysis, being unable to contribute in
understanding the existing trends and only making the plots more
  crowded without adding information.
After all selections, we end up with 122 data points sampling the
$z$-\ms-SFR grid (see Fig.~\ref{fig:ms3d}). Dust masses measured in
each bin of our grid are listed in Tables \ref{tab:stat1} to
\ref{tab:stat5}.

Our dust masses are in very good agreement with those computed by
\cite{magnelli13} with the same recipe.

In addition to using the \cite{draineli07} templates, we also fit our data
with a library extracted from the chemospectrophotometric model
GRASIL \citep{silva98},  tested to reproduce the small galaxy sample
of \cite{santini10}, and with a simple modified blackbody, which
assumes a single-temperature dust distribution. For
consistency with the \cite{draineli07} model, we build a modified blackbody
with emissivity index $\beta=2$ and absorption cross section per unit
dust mass at 240~\mic~of 5.17 cm$^2$/g
\citep{lidraine01,drainelee84}.  
We find that the simplified assumption of
single-temperature leads to dust masses which are lower by a factor of
$\sim$~1.5 compared to those obtained with the more realistic assumption of a
multi-temperature grain distribution \citep[in agreement with
previous studies,
e.g.,][]{santini10,magnelli12a,magnelli12b,dale12,magdis12}.  Indeed,
the attempt of reproducing the Wien side and at the same time the
Rayleigh-Jeans side of the modified blackbody spectrum has the effect of
overestimating the dust temperature and hence underestimating the dust
mass.  However, the recent work of \cite{bianchi13} ascribes such
disagreement to possible inconsistencies in the treatment of dust
emission properties between the two approaches.  The GRASIL 
  library
fits dust masses 
larger than the \cite{draineli07} templates
by a factor of 
1.5 on average. A direct comparison between the parameters
  assumed by the two models is not possible, since GRASIL computes
  dust emission by considering the physical properties of each single
  grain, instead of assuming an average emissivity. One reason for the discrepancy could be that the optically
  thin assumption of \cite{draineli07} is not always verified (even if
true, this would not affect our results, which would be simply
offset). 
 The GRASIL library adopted, however, has not been tested to work in
the absence of submillimeter data. 
Both the fit with GRASIL
and with the modified blackbody provide $\chi^2$ values that are a
factor of 1.5--2 larger than the \cite{draineli07} library. For these
reasons we decided to use the dust masses obtained from the
\cite{draineli07} templates. We will expand the GRASIL library by
enlarging the parameter space to better reproduce our galaxies in a
future analysis. In any case, we note that the effect of
choosing one dust model or the other only produces an offset, leaving
the main trends outlined below almost unchanged.

\subsection{Gas metallicities and gas masses}\label{sec:gmetmgas}

It is possible to take a further step forward 
with respect to observables
directly measurable from our data and compute 
gas masses by converting dust masses through the dust-to-gas ratio \citep[e.g.,][]{eales10}. In order to do that, we
  need to make some assumptions. 

We first
 assume that the gas metallicity is described by the 
 FMR
of \cite{mannucci10}. The FMR is a 3D relation between gas
metallicity\footnote{Gas metallicities were measured from
    emission line ratios following \cite{nagao06} and
    \cite{maiolino08}, i.e., from the [\ion{N}II]/H$\alpha$ ratio and/or from the
   R23$=$([\ion{O}II]+[\ion{O}III])/H$\beta$ quantity.}, stellar mass and SFR, with a very small scatter (0.05
dex). We assume that it does not evolve from the local Universe to $z\sim
2.5$, as confirmed by a number of recent works
\citep[e.g.][]{mannucci10,cresci12,nakajima12,henry13a,henry13b,belli13}.  
More recently, \cite{bothwell13} have shown that the FMR is
likely a by-product of a more fundamental relation, between \ion{H}I gas
mass, stellar mass and metallicity (\ion{H}I-FMR). However, it is beyond the
scope of the paper to discuss the origin of this relation.  
Given the average \ms~and SFR in each bin of our grid, following
\cite{mannucci10}, we
compute the gas metallicity  from the linear combination
  $\rm \mu_{0.32}=\log M_{star}-0.32 \log SFR$, 
after converting to a Chabrier IMF (as they adopt) both stellar
masses ($\log{\rm M_{star}^{Cha}}=\log{\rm M_{star}^{Sal}}-0.24$,
\citealt{santini12a}) and SFR ($\log{\rm SFR^{Cha}}=\log{\rm
  SFR^{Sal}}-0.15$, \citealt{dave08}),  using their equations 4 and 5 and the extrapolation for low $\mu_{0.32}$
  values published in \cite{mannucci11}. The inferred gas
  metallicities are in the range 8.58--9.07, with a scatter of
    0.14 dex around the mean value of $\sim 8.9$.

 We note that the FMR has not been tested over the entire SFR
  range studied in this work on large galaxy samples, so the extrapolation to SFRs larger than
  $\rm \sim 100 M_\odot/yr$ might in principle result in gas
  metallicity estimates that are incorrect. 
Moreover, the detailed shape of the FMR is matter of debate
  \citep[e.g.][]{yates12,andrews13}.  For these reasons, we also
  tested the robustness of our results by adopting the redshift-dependent
  mass-metallicity relations published by \cite{maiolino08} and verified
  that all our results are independent of the specific description of the
  gas metallicity. 

As suggested by previous studies,
focused either on  local
  \citep[e.g.,][]{draine07,leroy11,smith12,corbelli12,sandstrom13},
  $z<0.5$ \citep[e.g.,][]{james02} or high-$z$ galaxies 
  \citep[e.g.,][Cresci et al. in prep.]{zafar13,chen13},  
we consider that a fixed fraction of metals are
incorporated in dust. Within the metallicity range probed by our
  sample, this is true within 0.3 dex at most.   Following the parameterization
 provided by
\cite{draine07}, we assume
 that the dust-to-gas ratio ($\delta_{\rm DGR}$) scales  linearly with
the oxygen abundance through the constant factor $k_{\rm DGR}$:
\begin{eqnarray}
\delta_{\rm DGR} & = k_{\rm DGR}  \times (O/H)  = 0.01 \times (O/H)/(O/H)_{MW}=\nonumber\\\ 
& = 0.01 \times 10^{Z-Z_\odot},
\label{eq:dgr1}
\end{eqnarray}
where $Z=12+\log(O/H)$ is the gas metallicity and $Z_\odot = 8.69$ is
the Solar value \citep{allendeprieto01,asplund09}.  We find almost identical results  from our analysis if we
apply the linear relation between $\log \delta_{\rm DGR}$ and gas metallicity inferred by
\cite{leroy11}. 

 The universality of the depletion factor of metals into dust is
 outlined by the recent work of \cite{zafar13}. According to their analysis, the dust-to-metal
ratio can be considered universal, independent of either column
density, galaxy type or age, redshift and metallicity. However,
\cite{decia13} claim that the dust-to-metal ratio is
significantly reduced with decreasing gas metallicity at $Z<0.1
Z_\odot$ and low column densities. Yet, this should not be a concern for
our analysis, since our sample does not include such low-metallicity
galaxies. In a more recent paper, \cite{chen13}
combine constraints on the dust-to-gas ratio of lensed
galaxies, GRBs and quasar absorption systems, and find support for a simple,
linear universal relation between dust-to-gas ratio and
metallicity.

The total gas mass (atomic $+$ molecular, \mg~hereafter) can be computed as 
\begin{equation}
{\rm M_{gas}}= {\rm M_{dust}}/\delta_{\rm DGR}  .
\label{eq:dgr2}
\end{equation}

We can finally compute the gas fraction (\fgas~hereafter) as 
\begin{equation}
{\rm f_{gas}} = {\rm M_{gas}}/({\rm M_{gas}}+{\rm M_{star}}) .
\label{eq:fgas}
\end{equation}
The dust content, typically negligible with respect to the gas and
stellar mass components (\md~$\lesssim 0.01$ \ms, see below), is ignored
in the computation of \fgas. 

\section{Dust scaling relations}\label{sec:scaling}

In this section we investigate the correlations between \ms, SFR and
\md, and their evolution with redshift. 

\subsection{Dust content vs SFR} \label{sec:mdsfr}

\begin{figure}[!t]
\resizebox{\hsize}{!}{\includegraphics[angle=0]{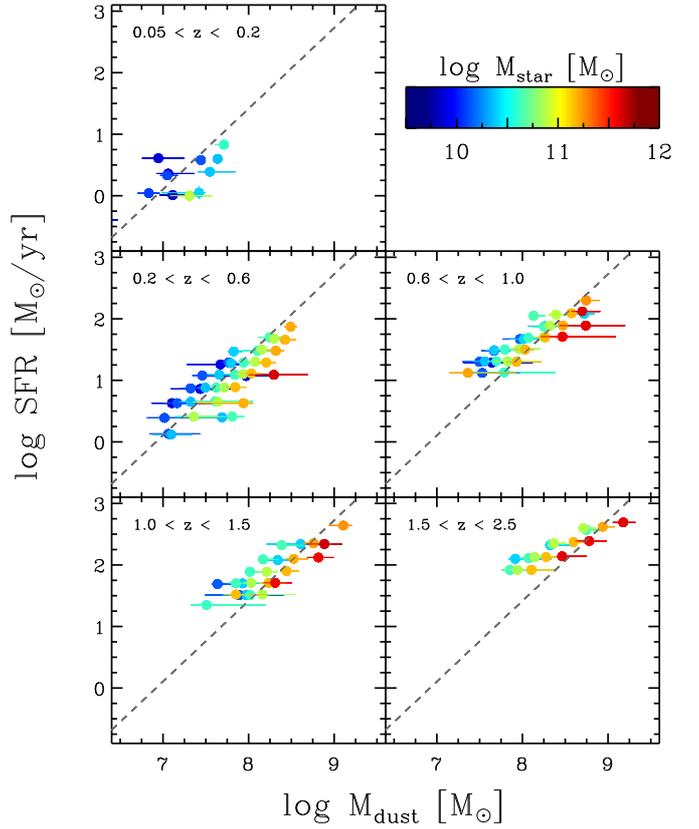}}
\caption{SFR vs dust mass in different redshift ranges. Galaxies are
  colour coded according to their stellar mass, as shown by the
  colour bar.  The dashed lines corresponds to the integrated
  Schmidt-Kennicutt law fitted by \cite{daddi10}, under the
    assumption of Solar metallicity (see text) and converted
  to a Salpeter IMF.  }
\label{fig:sfrmdust}
\end{figure}

\begin{figure}[!t]
\resizebox{\hsize}{!}{
\includegraphics[angle=0]{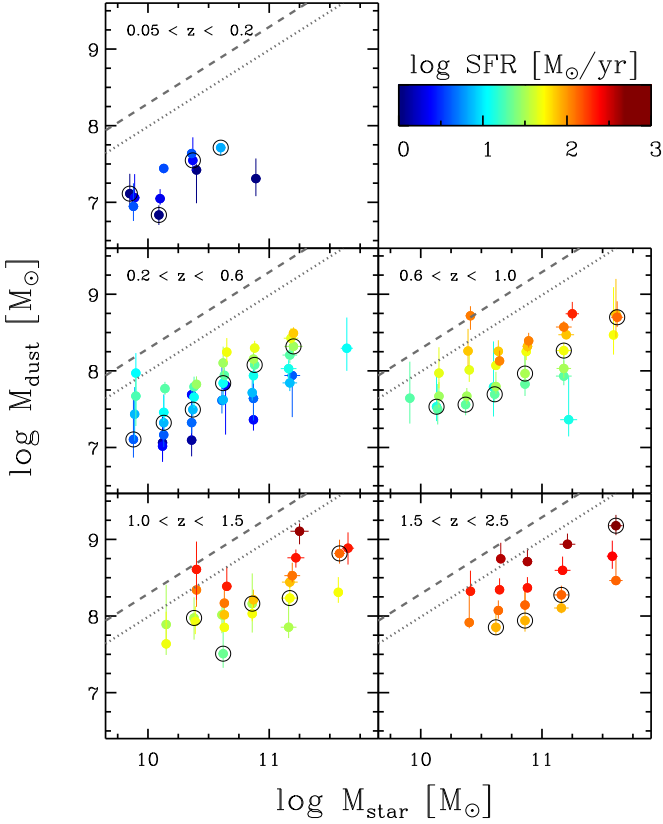}}
\caption{Dust mass vs stellar mass in different redshift
  ranges. Symbols are colour coded according to their SFR, as shown by
  the colour bar. 
 At each \ms, black open
  circles mark the bin which lies closest to the MS (in each \ms~interval), and in every case
  within 0.3 dex from it. The correlations between \md~and \ms~are
  rather flat when the data points are separated by means of their
  SFR.  The dashed lines correspond to an amount of dust equal to the maximum metal mass ${\rm
    M_Z}=y_{\rm Z} \times \rm{M_{star}}$, where $y_{\rm Z}\sim 0.014$,
  assuming the extreme case of a condensation efficiency of 100\%, while the dotted line shows
  the case when only 50\% of the metals are depleted into dust grains.
}
\label{fig:mstarmdust}
\end{figure}

Figure \ref{fig:sfrmdust} shows the relation between the SFR and the
dust content for galaxies of different \ms~at different redshifts. A
correlation between the dust content and the star formation activity
is evident at all \ms~and at all redshifts, although with some scatter, while no clear effect is
observed with the stellar mass, with bins of
  different \ms~sometimes overlapping (see also next section).

Before discussing the interpretation of this correlation, we
  stress here that, not only \md~and the SFR are estimated from
  different observed fluxes (Herschel and 24~\mic~bands, respectively)
  to avoid any possible degeneracy  and with intrinsically
    independent methods, but also they are not expected to be correlated by definition. 
  The SFR (although in our
  case measured from 24~\mic~observations) is in principle linked to
  the integrated IR luminosity, i.e., it is linked to the
  normalization of the far-IR spectrum.  The dust mass comes from a
  combination of the template normalization and temperature(s), which
  determines the shape; since the template library that we have used
  contains multiple heating source components, the dust mass is not
  trivially proportional to the SFR, though related to it through the
  dust temperature. To verify that any observed correlation is
  physical and not an obvious outcome of the relation between
  correlated variables, we run a simulation that is described in
  Appendix \ref{app:simul}, showing that, by starting from a
  completely random and uncorrelated distribution of dust masses and
  SFRs, our method does not introduce any artificial correlation.  

The
correlation observed in Fig.~\ref{fig:sfrmdust} primarily tells
  us that the dust temperature plays a secondary role. The SFR--\md~correlation
is clearly a
consequence of the S-K law, linking the SFR to the gas content.
Indeed, as shown in Sect. \ref{sec:gmetmgas}, the dust mass is 
related
to the gas mass by means of the dust-to-gas ratio.  In other
  words, from the S-K relation, we expect the dust mass to be 
roughly proportional to the gas mass, with
  the gas metallicity introducing minor effects through the
  dust-to-gas ratio.   
 Before converting dust masses into gas masses by adopting the
  appropriate dust-to-gas ratio in the next section, in order to represent  the S-K
 relation on a SFR vs \md~plot, for the moment we assume
 a constant dust-to-gas ratio for all galaxies. 
By using equation~\ref{eq:dgr2}, 
the S-K law (in
its integrated\footnote{ The term ``integrated'' refers to the
  measured power law relation between the gas mass and the SFR (see
  Sect. \ref{sec:sklaw}).}  version inferred by \citealt{daddi10} for
local spirals and $z\sim 2$ BzK galaxies, \citealt{daddi04})
 can be written 
 in terms
of SFR as a function of \md~as
\begin{equation}
\label{eq:daddi2}
\rm \log SFR[M_\odot/yr] = 1.31 \times \log
\left(\frac{M_{dust}[M_\odot]}{\delta_{\rm DGR \odot}} \right) +7.80 , 
\end{equation} 
where the last term includes the factor ($1.8\times 10^{-10}$) used to
convert  the total infrared luminosity 
(the original quantity  in the expression given in Daddi et al.) into
SFR, as well as the offset of 0.15 needed to convert from a Chabrier to a
Salpeter IMF (see Sect. \ref{sec:gmetmgas}),
 and $\delta_{\rm DGR \odot} $ is the dust-to-gas ratio computed from
  equation~\ref{eq:dgr1} by assuming a constant Solar metallicity. 
The dashed line in
Fig.~\ref{fig:sfrmdust} shows the inferred S-K relation on the
SFR--\md~diagram.

Our observational points follow reasonably well the trend expected
from the S-K law, with some scatter and a systematic trend 
  (flatter slope) at high-$z$. 
We will discuss this in Sect.~\ref{sec:sklaw}, where we also account for the variation
of the metallicity (hence the variation of the dust-to-gas ratio as a
function of metallicity).

\subsection{Dust vs stellar mass content} \label{sec:mdms}

\begin{figure}[!t]
\resizebox{\hsize}{!}{\includegraphics[angle=0]{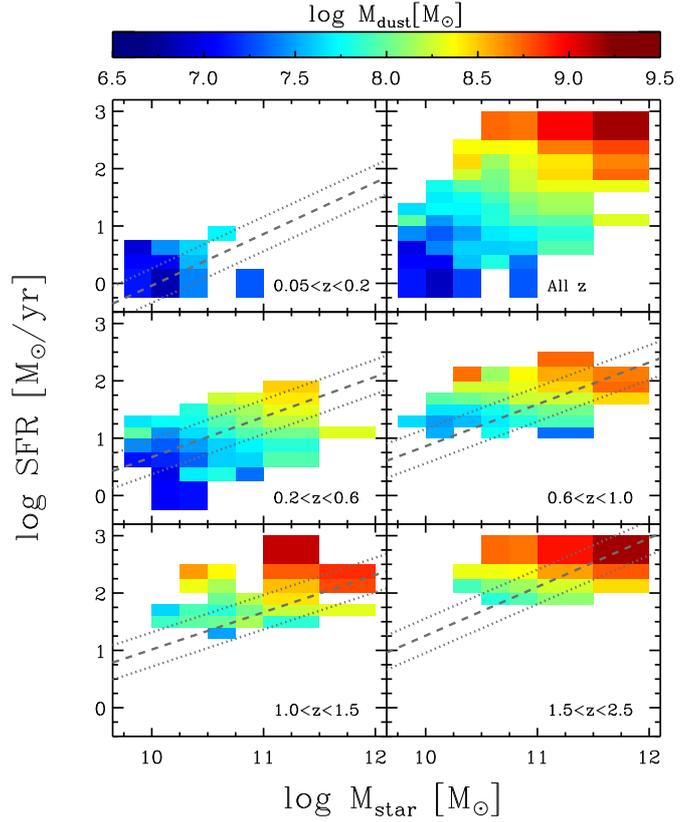}}
\caption{Average dust mass values, as indicated by the colour
  according to the colour bar, for bins of different SFR and
 \ms~in
  different redshift intervals and at all redshifts (upper right
  panel).  Dashed lines represent MS relations of star-forming
  galaxies as taken from the
 literature; the local MS is from
  \cite{peng10} (computed
 using \citealt{brinchmann04} data), rescaled
  to a Salpeter IMF, while
 the relations at higher redshifts are
  from \cite{santini09}. Dotted
 lines represent the $\pm 1\sigma$
  ($=0.3$ dex) scatter of the 
 MS relation.  }
\label{fig:ms3d}
\end{figure}

 We plot in Fig.~\ref{fig:mstarmdust} the dust mass as a function of
the stellar mass in bins of redshift. When the galaxies are separated
according to
 their SFR (coded with different colours), the
correlation found by previous authors
\citep[e.g.,   at low redshift by][]{bourne12} 
becomes much flatter and sometimes even disappears, hinting that this
correlation is at least partly an  indirect 
effect driven by other
phenomena.  More specifically, the \md--\ms~correlation is partly
 a
consequence of the \md--SFR correlation, reported in the
 previous
section, combined with the MS, i.e., the relation between SFR
 and
\ms. When all SFR are combined together, the low mass bins are
dominated by low SFR (as a consequence of the MS), which are
associated with low \md~(because of the SFR--\md~
relation). On the other hand, high mass bins are dominated by high SFR and
therefore associated with
 high \md. This results into an apparent
\ms--\md~correlation. To better visualize this effect in studies that
combine together all galaxies (i.e., without binning in a grid of SFR
and \md), in Fig.~\ref{fig:mstarmdust} we have
 marked with black
circles the bins closest to the MS (and in every case within
 0.3 dex from
it). These are
 the bins where the bulk of the star-forming galaxy
population is concentrated, and,
 as expected, they show a steeper
\ms--\md~trend compared to bins of constant SFR.

\begin{figure*}[!t]
\resizebox{\hsize}{!}{\includegraphics[angle=90]{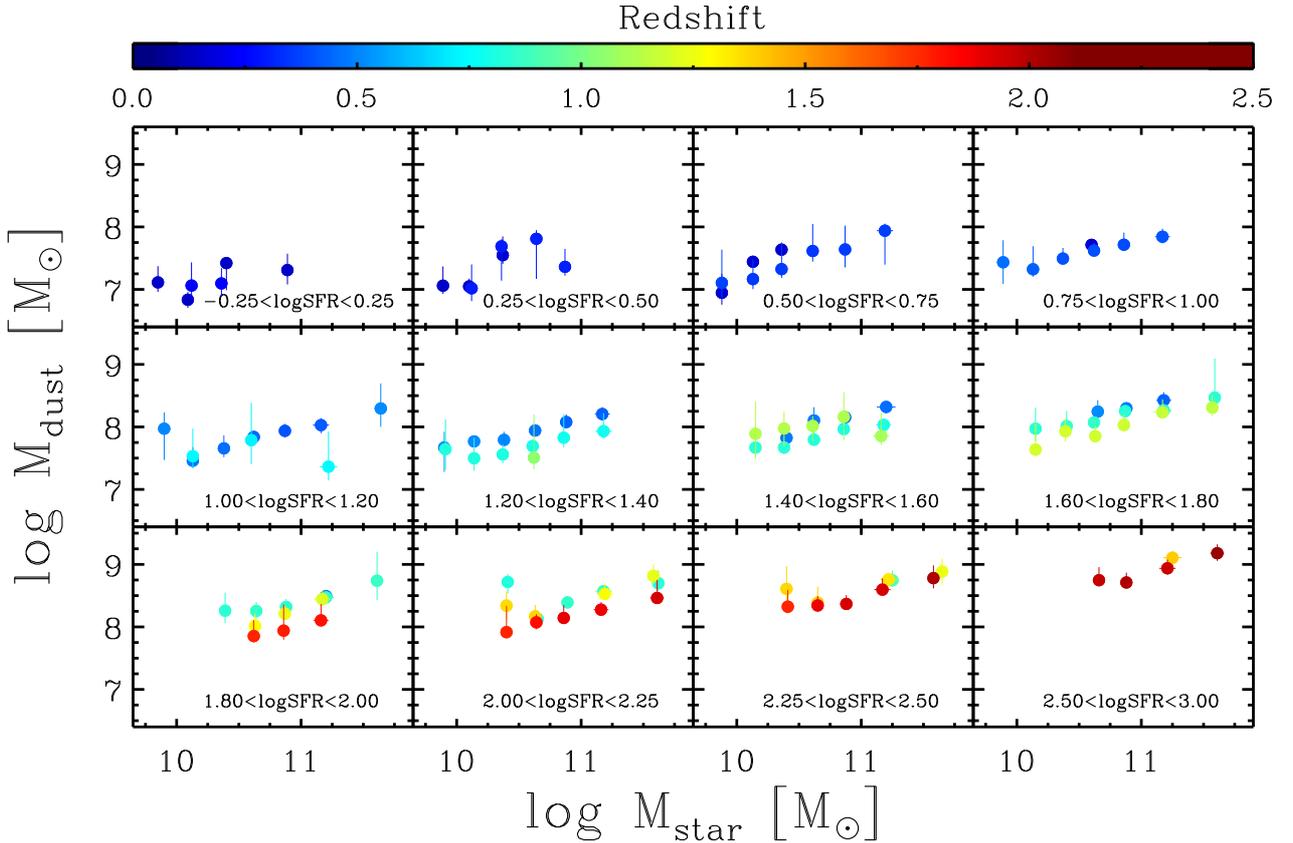}}
\caption{Dust mass vs stellar mass in panels of different SFR. The
  symbol colour indicates the mean redshift of each bin, as coded by
  the colour bar. No evolution with redshift is observed within
  uncertainties at a given \ms~and SFR.  }
\label{fig:zevol}
\end{figure*}

The dashed line in Fig.~\ref{fig:mstarmdust} represents the
 expected
maximum amount of metals (${\rm M_Z}=y_{\rm Z} \times
\rm{M ^{tot}_{star}}$, where $y_{\rm Z}\simeq 0.014$  and
$\rm{M ^{tot}_{star}}$ is the total stellar mass formed, including the
final products of stellar evolution\footnote{The fraction of stars
  which goes back into the ISM is  $\sim 30\%$ for a Salpeter IMF
  \citep{treu10}.}) produced by stars and
supernovae explosions, associated with the star formation required to
account for the observed \ms. This is also the maximum amount of dust
that can be associated with a given \ms~in a ``closed box'' scenario and
assuming a condensation efficiency
 in the ejecta close to 100\%.
More realistically, of these metals
 only about 30--50\% (\citealt{draine07}, 
grey dotted line in Fig.~\ref{fig:mstarmdust}) are expected to be
depleted into dust
 grains.  These lines give the maximum amount of
dust expected as a function of
 stellar mass if the galaxy behaves as
a ``closed box'', and metals
 are condensed in dust grains with
reasonable/high efficiency. 
 Most of the galaxies, in particular the
high mass systems, lie below
 the ``closed box'' lines. This finding
qualitatively agrees with the expectations of theoretical models for
the evolution of the dust content: 
 rather flat \md--\ms~trends,
i.e., decreasing dust-to-stellar mass
 ratios as the gas is consumed
and transformed into stars \citep[see, e.g.,][]{eales96,calura08,dunne11}.  Alternatively, this result might indicate that most of the
dust in these
 systems is  lost. In support of this scenario,
independently of the dust information, it has been
 acknowledged that
massive galaxies have a deficit of metals,
 by a factor of a few,
relative to what must have been produced in
 the same galaxies
\citep{zahid12}, which is ascribed to winds that
 have expelled
metal-rich gas out of these massive galaxies. On the
 contrary, hints
can be seen 
 for low \ms~galaxies (\msbin~$\lesssim
 9.75$) to
show a high dust mass, close to the maximum ``closed box'' limit.
Recent studies based on SPIRE data in the local and low-$z$
($z<0.5$) Universe support this evidence: large dust-to-stellar mass
ratios were reported
 by \cite{smith12}, while anti-correlations
between the dust-to-stellar
 mass ratio and stellar mass were
observed by \cite{cortese12} and
 \cite{bourne12}. 
 Due to the 
necessity of a careful check of optical counterpart associations to IR
galaxies with low \ms, we do not extend this work to such
 low
stellar masses. 
 The dust content in low \ms~galaxies will be
investigated by means of a dedicated analysis in a forthcoming paper.

\subsection{Summary view}

To give a global view of these correlations, we show in
Fig.~\ref{fig:ms3d} the SFR--\ms~plane at different redshifts, where
each bin is colour coded according to the associated dust mass.  We
also show  MS relations from the literature (from \citealt{peng10}
at $z\sim0$ and from \citealt{santini09} at high-$z$). 
This
representation gives a quick overview on the scaling relations
existing between \ms, SFR and \md: a weak and sometimes
absent trend of \md~with \ms~and a clear correlation between
\md~and SFR. 

It is also worth noting that we observe no evidence for evolution of
\md~across the different redshift ranges at a given \ms~$and$ SFR;
the main difference between the various redshift panels in
Fig.~\ref{fig:ms3d} is simply that they are populated differently.  To
make this more clear, Fig.~\ref{fig:zevol} shows \md~as a function of
\ms, in bins of SFR, where the colour coding identifies different
redshift bins (note that, as a by-product, Fig.~\ref{fig:zevol}
provides further evidence of weak/absent dependence of \md~on \ms~at a
fixed SFR).  At a given \ms~$and$ SFR, there is no clear evidence for
evolution of \md~with redshift within uncertainties.  We note,
however, that we cannot firmly exclude a decrease in \md~by a factor
of 2 from low- to high-$z$, though this trend is in a few cases
reversed. However, observational uncertainties on our data do not
allow us to claim any redshift evolution.

It is certainly true that, on average, the overall amount of dust in
galaxies at high redshift is higher, as a consequence of the overall
higher ISM content in the bulk of high-$z$ galaxies (see
Sect. \ref{sec:fgasms}).  As a matter of fact, the normalization of
the MS, representing the locus where the bulk of the population of
star-forming galaxies lies, does increase with redshift
\citep[e.g.,][]{santini09,rodighiero10,karim11} and, as a consequence,
the dominant galaxy population moves towards larger SFR, hence being
characterized by larger dust masses (Fig.~\ref{fig:sfrmdust}).
However, our results indicate that galaxies with the same properties
(same SFR $and$ same \ms) do not show any significant difference in
terms of dust content across the cosmic epochs, at least out to
$z\sim$2.5. In other words, dust mass in galaxies is entirely
determined by the SFR and, to a lesser extent, by \ms, and it is
independent of redshift within uncertainties. Put simply, different
cosmic epochs are populated by galaxies with different typical SFR and
\ms~values, and hence are characterized by different dust masses.

At fixed SFR, a non evolving \md~translates into a non evolving dust
temperature ({$\rm T_{dust}$}).  This does not contradict the results
of \cite{magnelli13}, presenting only a very smooth negative evolution
in the normalization of the {$\rm T_{dust}$}--specific SFR
(SSFR$=$SFR/\ms) relation. They also find a stronger positive
evolution in the normalization of the relation between {$\rm
  T_{dust}$} and the distance from the MS. However, as discussed
above, the normalization of the MS itself increases with redshift,
hence different SFR--\ms~combinations are probed at different epochs.

Given the lack of any significant redshift evolution in the dust mass
at fixed \ms~$and$ SFR, it is meaningful to represent all redshift
bins on the same SFR--\ms~plane (upper right panel of
Fig.~\ref{fig:ms3d}) to provide an overview of the dust content over a
wider range of \ms~and SFR.  Here the dust mass is computed by
  averaging the values at different redshifts. This further confirms the trends already
outlined (\md~depends strongly on the SFR and weakly on \ms), over a
wider dynamic range.

\section{The evolution of the gas content in galaxies} \label{sec:gascontent}

We investigate here the relation between the gas content and the SFR,
as well as the evolution of the gas fraction, with the aim of
understanding the processes driving the conversion of gas into stars
in
 galaxies throughout the cosmic epochs. We recall that gas masses
are inferred
 from dust mass measurements by assuming that the
dust-to-gas ratio
 scales with the gas metallicity, and by computing
the latter by means of
 the FMR of \cite{mannucci10} (see equations
\ref{eq:dgr1}--\ref{eq:dgr2}). We verified that all the results
presented below are almost unchanged if the redshift-dependent
mass-metallicity relation of \cite{maiolino08} is used instead of the
FMR.

\subsection{The star formation law} \label{sec:sklaw}

\begin{figure*}[!t]
\resizebox{\hsize}{!}{\includegraphics[angle=90]{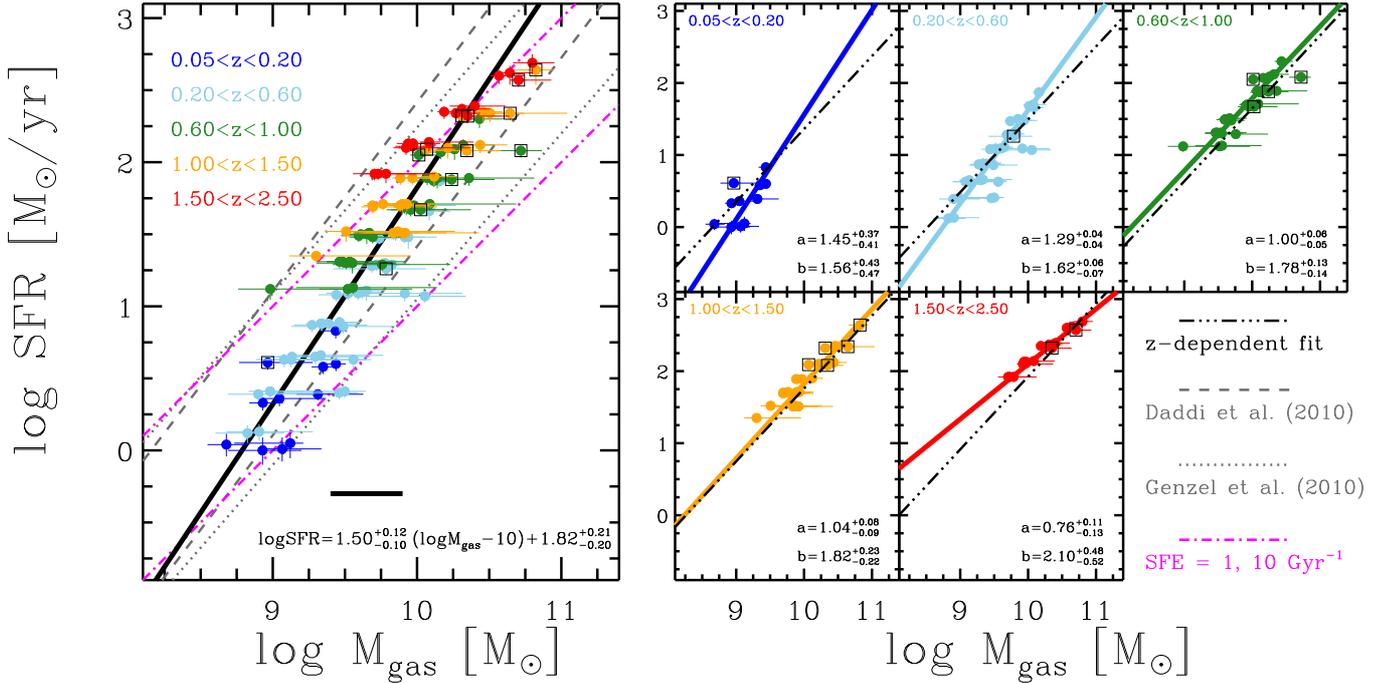}}
\caption{{\it Left panel: } Relation between SFR and gas mass. The
  colour code indicates different redshift intervals, as shown by the
  legend in the upper left corner. The black boxes mark bins that lie
  in the starburst region according to \cite{rodighiero11}. The solid
  thick black line is the power law fit to all data, and the best-fit
  relation is reported in the lower right corner.  The dashed and
  dotted grey lines show the integrated Schmidt-Kennicutt relation
  fitted by
 \cite{daddi10} and \cite{genzel10}, respectively, on
  normal star-forming
 galaxies (lower curves) and on local ULIRGs
  and $z\sim 2$ SMGs (upper
 curves).  Curves from the literature are
  converted to a Salpeter IMF. Magenta dashed-dotted lines indicate
  constant star formation efficiencies (i.e., constant depletion
    times) of 1 (lower curve) and 10 (upper curve) Gyr$^{-1}$. {\it
    Right panels:} Relation between SFR and gas mass in different
  redshift bins, indicated in the upper left corner of each
  panel. Symbol styles and colours are as in the left panel. The
  coloured solid curves are the power law fits to the data, and the
  numbers in the lower right corner indicate the best-fit slope (upper)
  and intersection at $\rm \log M_{gas}[M_\odot]=10$ (lower) (see equation~\ref{eq:sklaw}). The
  dashed-triple dotted lines show the best-fit relation given in
  equation~\ref{eq:sklaw_zdep} calculated at the median redshift in
  each panel.  }
\label{fig:sklaw}
\end{figure*}

We plot in Fig.~\ref{fig:sklaw} the values of SFR as a function of gas
mass. The colour code identifies bins of different redshift.  For the
sake of clarity, the data points at the different redshifts are also
plotted on separate panels on the right side. This figure is analogous
to Fig.~\ref{fig:sfrmdust}, except that  \mg, plotted here instead
of \md, 
takes into
account the dependence of the gas metallicity with stellar mass and
SFR  (see Sect. \ref{sec:gmetmgas}). This, however, introduces only a
minor effect (the gas metallicity changes less than a factor of 2--3 or less,
 while the dust mass spans 2--3 orders of magnitude). The
relation shown in Fig.~\ref{fig:sklaw} can be referred to as the
integrated S-K law, meaning that gas masses and SFRs are investigated
 values 
rather than their surface
 densities, as in the original S-K law,
where the SFR surface density is
 related to the gas surface density
by a power law
 relation. We fit 
the data points with the relation
\begin{equation}
\log {\rm SFR} = a ~(\log {\rm M_{gas}-10})   + b .
\label{eq:sklaw}
\end{equation}
A standard $\chi^2$ fit cannot be performed on our data given the
asymmetric error bars. Therefore, all over our work, we apply a
maximum likelihood analysis by assuming rescaled log-normal shapes for
the probability distribution functions of the variables with the
largest uncertainties ($\rm \log M_{gas}$ in this case) and by
ignoring the uncertainties on the other variables.  By fitting the
total sample we obtain $a=1.50^{+0.12}_{-0.10}$ and
$b=1.82^{+0.21}_{-0.20}$, where a bootstrap is performed to compute
the parameter 1$\sigma$ errors.  The best-fit relation is represented
by the black solid line in the left panel of
Fig.~\ref{fig:sklaw}. However, due to inhomogeneous sampling in SFR at
different redshifts, the fit might suffer from biases in case there is
an evolution in the slope or normalization of the relation. To
investigate such effects, we also separately fit the points in each
individual redshift bin (coloured solid lines in the right panels of
Fig.~\ref{fig:sklaw}). The inferred slopes monotonically decrease with
redshift from $1.45^{+0.37}_{-0.41}$ in the local Universe to
$0.76^{+0.11}_{-0.13}$ at $z\sim2$, while the normalizations increase
from $1.55^{+0.43}_{-0.47}$ to $2.10^{+0.48}_{-0.52}$.  
The best-fit parameters are given in the bottom right corners of each
panel of Fig.~\ref{fig:sklaw}. 

By following the theoretical model of \cite{dave11,dave12} and
  the observational results of \cite{tacconi13}, we also attempt to
  fit 
  our data points with a relation that has a single
  redshift-independent slope and normalization slowly evolving with
  redshift, i.e., yielding a cosmological scaling of the depletion time
  ($=$\mg/SFR):
\begin{equation}
\log {\rm SFR} = m ~(\log {\rm M_{gas}-10})   + n ~\log (1+z) + q .
\label{eq:sklaw_zdep}
\end{equation}
The best-fit parameters are $m=1.01^{+0.14}_{-0.17}$,
$n=1.40^{+0.85}_{-0.74}$ and  $q=1.28^{+0.14}_{-0.17}$. 
The dashed-triple dotted lines in the right panels
of Fig.~\ref{fig:sklaw} show the inferred relation at the median
redshift in each bin.  This function provides 
a worse fit to the data in terms of  probability of
  the solution as computed from the likelihood, with respect to
  equation \ref{eq:sklaw}. 

In both cases, the evolution of the relation with redshift
may be partly caused by mixing different stellar masses, whose
contribution strongly depends on the SFR and redshift because of the
evolution of the MS relation.

\subsubsection{Comparison with previous works}\label{sec:comparison}

The inferred relations agree, on average, 
well with those fitted by previous work based on CO measurements for
normal star-forming galaxies (\citealt{daddi10}, lower dashed grey
line in Fig.~\ref{fig:sklaw}, see also equation~\ref{eq:daddi2}, and
\cite{genzel10}\footnote{We used the best-fit relation between FIR and
  CO luminosities in their figure 2 and the conversions given in their
  table 1 to convert to SFR and \mg, respectively.}, lower dotted grey
line), although we fit a steeper slope on all data points.  The values of the
  best-fit slopes are independent  from the galaxy population (i.e.,
  consistent fits are found if starburst galaxies are removed, see
  below), and of the  recipe adopted for the gas metallicity 
  (i.e., consistent results are obtained if we assume no dependence on the SFR and redshift evolution of the
  mass-metallicity relation). Anyhow, the broad
agreement with previous studies for the majority of galaxies (see
  below) and the small dispersion (the average absolute residual is
$\sim$~0.15
dex in terms of log \mg) shown by our
data points are impressive, especially given the completely different
and independent approaches used to derive
 the star formation law.
This confirms the
 reliability of our approach of deriving gas mass
estimates from dust
 mass measurements. 

We remind the reader that the dust method is supposed to trace both
the molecular and atomic gas (the dust-to-gas conversion factor
adopted refers to the total gas mass).  \cite{bigiel08} have measured
steeper slopes for the star formation laws in local galaxies when both
the molecular and atomic gas components are considered. This may
explain our steeper slopes compared to previous CO-based studies
\citep[e.g.,][]{genzel10,tacconi13}.  However, the fair agreement with
the \cite{daddi10} relation (inferred from CO observations, a proxy
for molecular hydrogen) is suggesting that, if the latter is correct,
the bulk of the gas in these galaxies is in the molecular phase, which
is reasonable given that most of these galaxies are vigorously forming
stars and will have high pressure ISM conditions (see also
\citealt{leroy09,magdis12}).  The steeper slopes found at low redshift
may be determined by a larger atomic-to-molecular gas ratio at low
than at high-$z$ (see below). Another possibility to explain this is
the trend for the MS template of \cite{elbaz11} to slightly
underpredict the SFR in the absence of Herschel data for bright
galaxies at high-$z$ \citep[SFR $\rm >100~ M_\odot/yr$, see
Fig.~\ref{fig:cfrsfr} and][]{berta13}; by moving the data points with
the largest SFR towards lower SFR values, this effect might be
responsible for the shallower slope measured at high
redshift. However, as it can be seen in Fig.~\ref{fig:cfrsfr}, this
effect is not larger than 0.1--0.2 dex, and is therefore unlikely to
affect our other results (on the other side, the fitted slope of the
S-K law may be sensitive to small offsets in the SFR). As a matter of
fact, the results presented in this paper are very similar if other IR
templates \citep[e.g.,][]{dh02} are used to measure the SFR from
24~\mic~fluxes or from all Herschel bands.  Finally, steep slopes for
the global star formation law may be explained by the results of
\cite{saintonge13}, who claim that the gas-to-dust ratio may be 1.7
times larger at $z>2$ than observed locally. This, however, would only
marginally affect our highest redshift bins, whose mean redshift value
is around 2.

Note that the fact that the slope of the global S-K relation, as well
as those at $z<0.6$, are steeper than unity implies that galaxies with
high star formation rates have higher star formation efficiency
(defined as SFE=SFR/\mg, equal to the inverse of the depletion time),
even if they are regular, MS galaxies.  The magenta dash-dotted lines
in Fig.~\ref{fig:sklaw} trace the loci with SFE~=~1~Gyr$^{-1}$ (lower
line) and SFE~=~10~Gyr$^{-1}$ (upper line). As a consequence of the
super-linear slope of the S-K relation, moderate star-forming galaxies
(SFR $\sim$ 1~$\rm M_{\odot}/yr$) have a SFE approaching 1~Gyr$^{-1}$,
while strongly star-forming galaxies (SFR~$\sim$~several 100~$\rm
M_{\odot}/yr$) have a SFE approaching 10~Gyr$^{-1}$, implying gas
depletion timescale of a few times 100~Myr.  However, the SFE is more
properly defined as the ratio of the SFR over the molecular gas
content. Therefore, another possibility to interpret our result is
that the SFR/$\rm M_{mol_{gas}}$ stays the same, and the atomic gas
content decreases in strongly star-forming galaxies, or, in other
words, the latter have a larger molecular to atomic fraction.  This
would be confirmed by the results of \cite{bauermeister10}, who
observe little evolution in the cosmic \ion{H}I density, while the
molecular component is expected to positively evolve out to the peak
of cosmic star formation ($z\sim$ 2--3,
\citealt{obreschkow09,lagos11,popping13}).

\subsubsection{The star formation law for starburst galaxies}
 
  Symbols marked with a black box in Fig.~\ref{fig:sklaw} correspond
  to
 bins which lie in the starburst region of the SFR vs
  \ms~diagram according to the 
  selection of \cite{rodighiero11}.  They select starburst galaxies as
  sources deviating from a Gaussian logarithmic distribution of the
  SSFR, having  SSFR  
four times higher than the peak of the distribution (associated
  to MS galaxies). 
  Given the average scatter of 0.3 dex of the MS
  \citep{noeske07}, these galaxies are located $>2\sigma$ above the MS
   (see Fig.~\ref{fig:ms3d}).  The effectiveness of this SSFR
    criterion in selecting starburst galaxies is confirmed by
  semi-analytical models where starburst events are triggered by
  galaxy interactions during their merging histories
  \citep{lamastra13}. Galaxies from our sample located in the
  starburst regions do seem to follow the same star formation law as
  all other galaxies.  We note that  the selection of starburst
  galaxies above is based on the knowledge of the MS from the
  literature, rather than computed directly on the present
  sample. However, this does not affect our conclusions. Indeed, the
  observed correlation between SFR and \mg~is tight enough that,
  even in case of small variations in the location of the MS, sources
  selected as starburst would still follow the same relation (for
    example, results are unchanged if the MS from \citealt{whitaker12}
    is used, despite its shallower slope). 
  Unless indicative of a larger fraction of atomic gas in starbursts,
  this result is in contrast with what suggested by previous studies,
  mostly based on CO emission
  \citep[e.g.,][]{daddi10,genzel10,saintonge12,magdis12,sargent13}.
  The latter studies find a normalization of the star formation
    law $\sim$~10 times higher for starburst galaxies, implying a
    larger SFE.  In any case, since the slope of the relation 
    that we infer is larger than unity (except at $z>0.6$),  our
    result does not imply a low efficiency in converting gas into
  stars for galaxies located in the starburst region (see
 next section):
  starburst galaxies do have, on average, larger star formation
  efficiency  (i.e., shorter depletion times) than the bulk of
  star-forming 
  galaxies (typically at lower SFR).

 We note that our work
  does not sample the most extreme objects lying at the bright tail of
  the SFR distribution (all but one of the bins selected as
    ``starbursts'' are located between 2$\sigma$ and 3$\sigma$ above
    the MS). Physical properties of very extreme 
  sources,  such as local ULIRGs or high-$z$ SMGs, are not always
  compliant with local-based expectations \citep[see,
  e.g.,][]{santini10} and need to be treated with ad-hoc techniques
  (for example, \citealt{magdis12} claim the need of submm data
    to reliably estimate dust masses of SMGs). Moreover, larger
  statistics is needed. We will therefore study such extreme sources
  in a future work.

\subsection{The evolution of the star formation efficiency}
 
\begin{figure}[!t]
  \resizebox{\hsize}{!}{\includegraphics[angle=0]{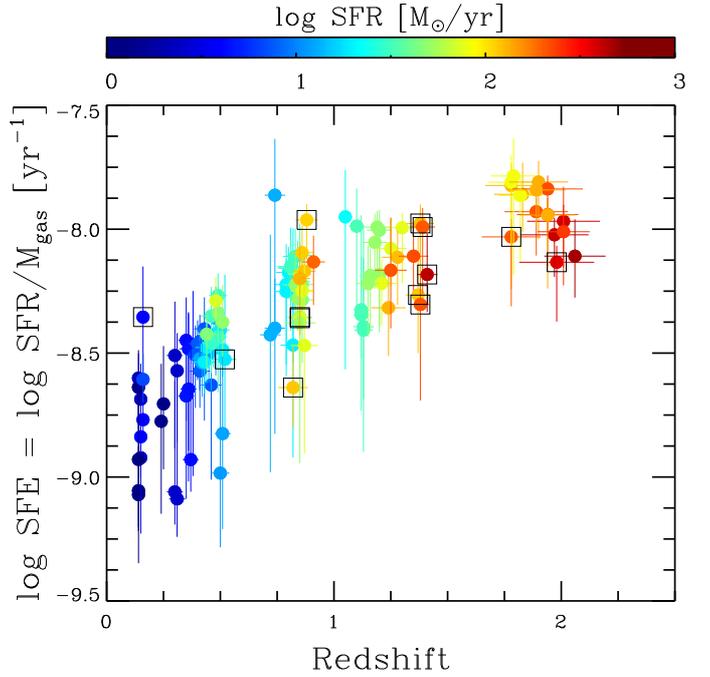}}
\caption{Redshift evolution of the star formation efficiency (SFE,
  or inverse of the depletion time). Different colours refer to
  different SFRs, as shown by the colour bar. Black boxes are as in
  Fig.~\ref{fig:sklaw}. 
}
\label{fig:sfe}
\end{figure}

The slope of the integrated S-K relation inferred from our data is
generally steeper than unity 
(except possibly at high redshift). As a consequence, the SFE for high
redshift galaxies, which are also on average more star-forming,
is higher than for local galaxies, or, equivalently, the depletion time
  is shorter (we here assume negligible atomic
fraction for all galaxies, but see comment in
Sect. \ref{sec:comparison}). This is illustrated in Fig.~\ref{fig:sfe},
where the SFE is plotted as a function of redshift, and where an
increase in the SFE with redshift is indeed observed, although with
large scatter. Due to degeneracy between SFR evolution and redshift it
is not clear whether the increase in the SFE with redshift truly
reflects a cosmic evolution of the SFE, i.e., galaxies of a given SFR
convert their gas into stars more efficiently at high-$z$, or it is
simply a by-product of the slope of the S-K relation convolved with the
higher SFR characterizing high-$z$ galaxies (higher normalization of
the MS). In Fig.~\ref{fig:sfe} galaxies with different SFRs are
plotted with different colours, in an attempt to break the degeneracy
between redshift and SFR.  Galaxies with similar SFR show no clear
internal evolution with redshift.  However, due to observational
  biases (difficulties in observing faint sources at high-$z$ as well
  as paucity of rare bright sources in small volumes at low-$z$) the
redshift spanned by each of these sets of points is very narrow, and
the dispersion very high, hence we cannot rule out a real, intrinsic
evolution of the SFE in galaxies (at a given SFR).


In any case, regardless of whether the evolution of the SFE is an 
  intrinsic redshift evolution or 
driven by the slope of the S-K relation and the evolution of the SFR,
the net result is that the bulk of the galaxy population (i.e., galaxies on the MS) at high
redshift ($z\sim 2$) do form stars with a SFE 
higher  by a factor of $\sim5$ than the bulk of the population of local star-forming
galaxies. 
This
  evolution is  roughly consistent with the evolution of the dust
    mass-weighted luminosity ($\rm L_{IR}/$\md, proportional
    to the SFE except for a metallicity correction) found by
    \cite{magdis12} (a factor of $\sim4$ from $z\sim0$ to $z\sim2$)
    and only slightly steeper than the evolution of the depletion time
    (a factor of $\sim3$ in the same redshift range)
observed by \cite{tacconi13}, 
  likely due to the steeper S-K law inferred by us
  compared to their work.

\subsection{The evolution of the gas fraction} \label{sec:nonevol}

 \begin{figure*}[!t]
\resizebox{\hsize}{!}{\includegraphics[angle=90]{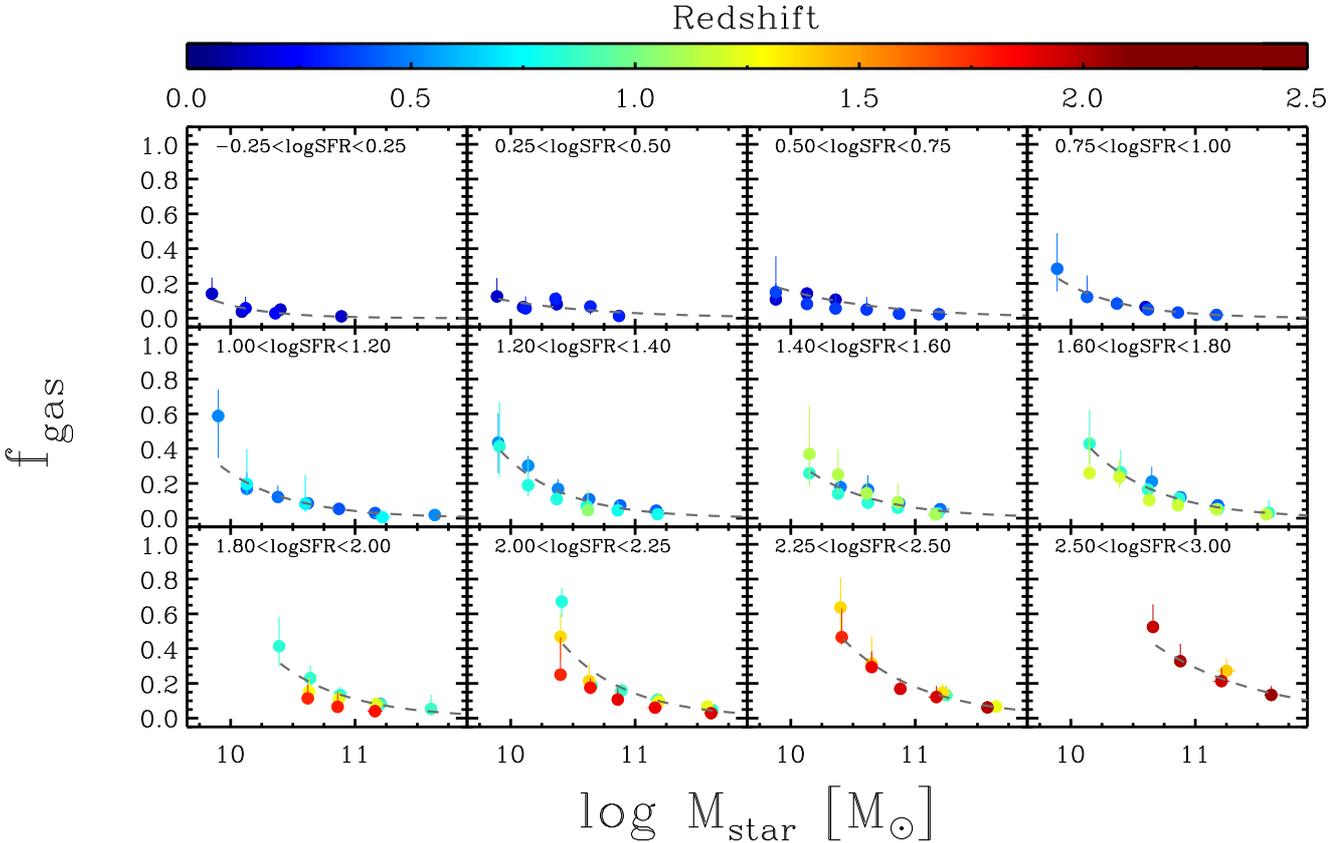}}
\caption{Gas fraction vs stellar mass in panels of different average
  SFR. The colour of the symbols reflects the mean redshift of each bin, as
  indicated by the colour bar. No evolution with redshift is observed,
  within uncertainties, at given SFR and \ms. Grey dashed curves are
  the best-fits to the data assuming the functional shape in
  equation~\ref{eq:fgasparam}. Best-fit parameters for each SFR interval
  are summarized in Table \ref{tab:param}.}
\label{fig:fgas_noevol}
\end{figure*}

Fig.~\ref{fig:fgas_noevol} shows the gas fraction as a
function
 of the stellar mass, colour coded according to the
redshift, in panels
 of different SFR.  The gas fraction decreases
with the stellar
 mass, as expected by the
gas conversion into stars in a closed-box model, 
 and increases with the SFR, as a
consequence of the S-K relation (see also the results of
\citealt{magdis12} and those of the PHIBSS
survey presented in \citealt{tacconi13}).  Most interesting is the lack of
evolution of the gas fraction with
 redshift, once galaxies are
separated according to their \ms~and
 SFR values.  Given the assumptions
made to compute the gas mass, hence gas
 fractions, this finding is
the result of the lack of (or marginal) evolution of the dust
 content in
bins of fixed \ms~{\it and} SFR (see Fig.~\ref{fig:zevol}), combined with
a
 minor contribution from the gas metallicity evolution with \ms~
and
 SFR (the FMR, \citealt{mannucci10}).  

From the lack of redshift evolution of the gas fraction at fixed SFR
{\it and} \ms, it follows that galaxies within a given population
(identified by a combinations of SFR and \ms), convert gas at the same
rate regardless of redshift, i.e., the physics of galaxy formation is
independent of redshift, at least out to the epochs probed by our
work. This is essentially a consequence of the unimodal inferred S-K
relation, but Fig.~\ref{fig:fgas_noevol} shows the result more neatly
by also slicing the relation through the dependence on stellar mass,
which is the third fundamental parameter.  We note that this does
  not contradict the evolution of the SFE observed in
  Fig.~\ref{fig:sfe}, where different stellar masses and SFR are mixed
  together and where selection effects cause the different SFR bins to
  be populated differently at different redshifts (hence the average
  at each redshift is certainly biased).

\begin{figure}[!t]
\resizebox{\hsize}{!}{\includegraphics[angle=0]{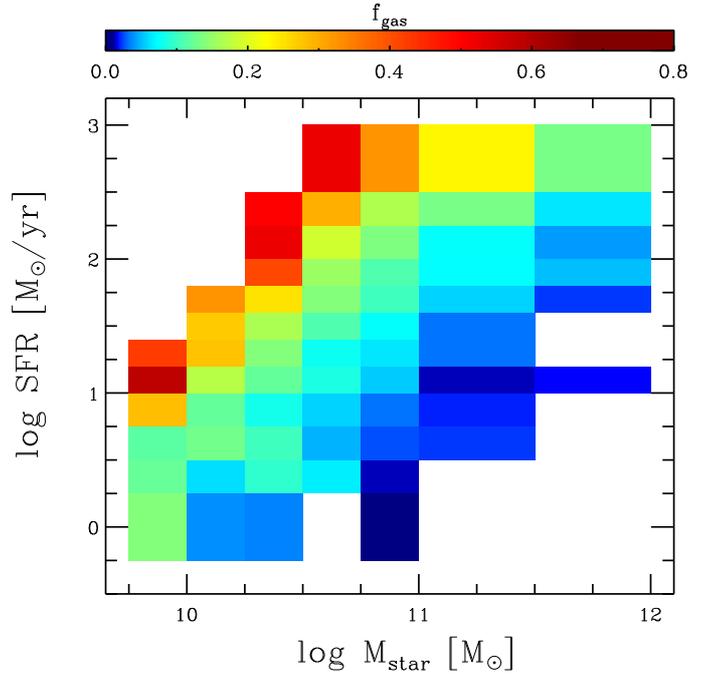}}
\caption{Average gas fractions, as indicated by the upper colour
    bar, in bins of \ms~and SFR. 
}
\label{fig:fgas_mssfr}
\end{figure}

In summary, our result implies that, at fixed stellar mass, the  SFR 
is uniquely driven by  the gas fraction via the star formation law.  In other
words, if two among SFR, \ms~ and \mg~are known, the third property
is completely determined and does not depend on redshift. This
provides a powerful tool to overcome the observational difficulties
related with the measurement of gas or dust masses and analyse the gas
content for much larger samples of galaxies.

\begin{figure}[!t]
\resizebox{\hsize}{!}{\includegraphics[angle=0]{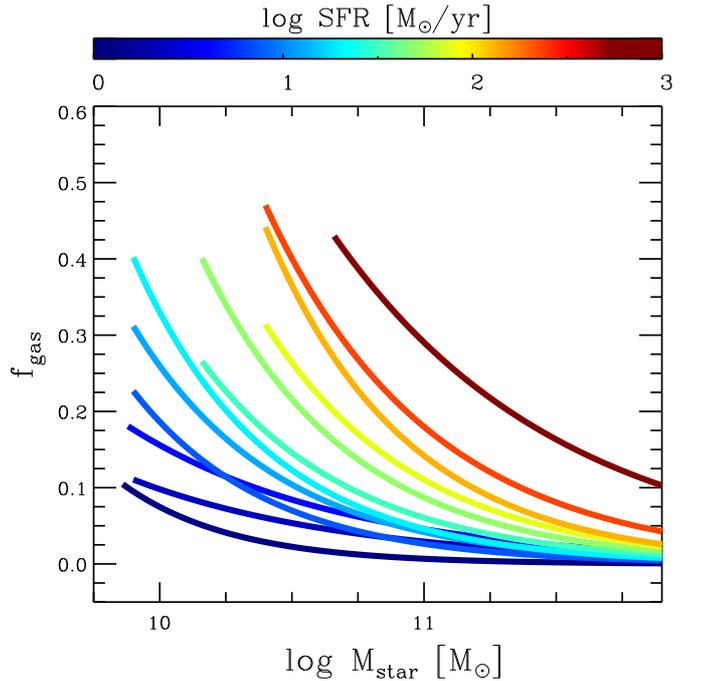}}
\caption{ Parameterization of the  gas fraction as a function of stellar mass at all redshifts in different SFR
  intervals, using  the functional shape given in
  equation~\ref{eq:fgasparam} (see text). Curves of different colours refer to different SFR bins, as
  shown by the colour bar. 
}
\label{fig:fgas_fit}
\end{figure}

\subsection{The {\it fundamental $f_{gas}$--$M_{star}$--SFR relation}}

Given the lack of evolution with redshift observed for the gas
fraction once galaxies with the same \ms~{\it and} SFR are considered, we
can combine all redshift bins together to increase the statistics and
infer more clearly the trend of \fgas~as a function of \ms~in
different SFR intervals.  Figure~\ref{fig:fgas_mssfr} shows the
resulting global dependence of the gas fraction (given by the colour
coding) on the SFR--\ms~plane.  In each SFR interval, we fit the data
points with a linear relation in the logarithmic space :
\begin{equation}
\rm \log  f_{gas}=\alpha+\beta (\log M_{star}-11) .
\label{eq:fgasparam}
\end{equation}
 We shift the stellar masses, placing them across zero, in order
  to de-correlate the slope and offset parameters in the linear fit
  result. 
The best-fit parameters 
are given in Table \ref{tab:param}, and the best-fit curves are
shown by the dashed grey lines  in Fig.~\ref{fig:fgas_noevol} and also by
the solid coloured lines in Fig.~\ref{fig:fgas_fit}, which provides
a direct comparison at different SFRs. 
We 
note that the functional form adopted above does not  necessarily have
physical meaning: it is
a purely phenomenological representation of the data to better visualize
the observed trends and to interpolate the three physical quantities for later
use of this 3D relation.  

\begin{table}
\centering
\caption{Best-fit parameters of the functional shape in
  equation~\ref{eq:fgasparam} describing the gas fraction as a function of the stellar mass in different SFR intervals.}
\begin{tabular} {cccc}
\hline 
\noalign{\smallskip} 
\sfrbin & $\alpha$ & $\beta$ & $\rm \log M_{star~min}$ \\
 \noalign{\smallskip} \hline \noalign{\smallskip}
$-0.25$ -- $0.25$ & $-2.17^{+0.16}_{-0.31} $ & $-1.04^{+0.32}_{-0.37} $ & $9.85
$ \\
\noalign{\smallskip}
$0.25$ -- $0.50$ & $-1.53^{+0.33}_{-0.35} $ & $-0.52^{+0.39}_{-0.39} $ & $9.89
$ \\
\noalign{\smallskip}
$0.50$ -- $0.75$ & $-1.34^{+0.14}_{-0.19} $ & $-0.53^{+0.20}_{-0.25} $ & $9.88
$ \\
\noalign{\smallskip}
$0.75$ -- $1.00$ & $-1.58^{+0.02}_{-0.02} $ & $-0.85^{+0.04}_{-0.05} $ & $9.89
$ \\
\noalign{\smallskip}
$1.00$ -- $1.20$ & $-1.38^{+0.03}_{-0.02} $ & $-0.79^{+0.09}_{-0.10} $ & $9.90
$ \\
\noalign{\smallskip}
$1.20$ -- $1.40$ & $-1.34^{+0.05}_{-0.05} $ & $-0.86^{+0.08}_{-0.08} $ & $9.90
$ \\
\noalign{\smallskip}
$1.40$ -- $1.60$ & $-1.22^{+0.05}_{-0.05} $ & $-0.77^{+0.10}_{-0.09} $ & $10.15
$ \\
\noalign{\smallskip}
$1.60$ -- $1.80$ & $-1.06^{+0.03}_{-0.03} $ & $-0.79^{+0.05}_{-0.08} $ & $10.15
$ \\
\noalign{\smallskip}
$1.80$ -- $2.00$ & $-0.96^{+0.02}_{-0.02} $ & $-0.76^{+0.11}_{-0.12} $ & $10.39
$ \\
\noalign{\smallskip}
$2.00$ -- $2.25$ & $-0.85^{+0.06}_{-0.05} $ & $-0.82^{+0.18}_{-0.15} $ & $10.40
$ \\
\noalign{\smallskip}
$2.25$ -- $2.50$ & $-0.75^{+0.06}_{-0.02} $ & $-0.70^{+0.07}_{-0.18} $ & $10.40
$ \\
\noalign{\smallskip}
$2.50$ -- $3.00$ & $-0.54^{+0.05}_{-0.03} $ & $-0.50^{+0.02}_{-0.15} $ & $10.66
$ \\
\noalign{\smallskip}
\noalign{\smallskip} \hline \noalign{\smallskip}
\end{tabular}
\tablefoot{The last column reports the minimum stellar mass sampled in
  each SFR bin. These parameterizations should not be employed below
  these limits.} 
\label{tab:param}
\end{table}

\begin{figure}[!t]    
 \resizebox{\hsize}{!}{\includegraphics[angle=0]{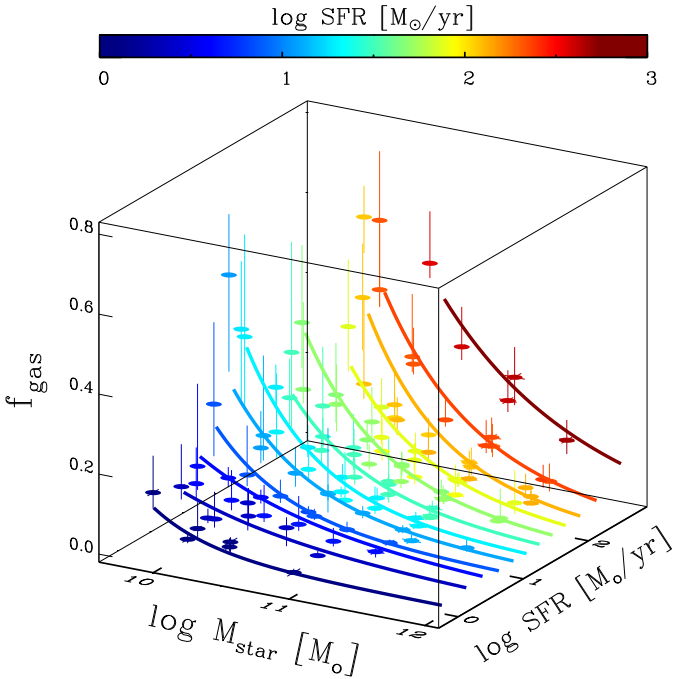}}
\caption{ Representation of the 3D {\it fundamental
    $f_{gas}$--$M_{star}$--SFR relation}. The colour code indicates
  the average SFR of each bin. The best-fit relations shown in
    Fig.~\ref{fig:fgas_noevol} are overplotted. 
  }
\label{fig:3d}
\end{figure}


The three-dimensional \fgas --\ms --SFR relation shown in
Fig.~\ref{fig:fgas_fit} is a {\it fundamental relation} that does not
evolve with redshift, at least out to $z\sim2.5$.  Galaxies move 
over this surface during their evolution. 

Fig.~\ref{fig:3d} shows a 3D representation of such a relation.
Further investigation of this 3D relation and its physical
interpretation goes beyond the scope of this paper, and will be
discussed in a future work, as well as the relation between the
  independent quantities \mg, \ms~and SFR. Here we only emphasize
that the redshift evolution of the S-K  law investigated in
  equation~\ref{eq:sklaw_zdep} seems to disappear once sources are
divided in bins of \ms.   Indeed, the redshift evolution of the
SFE illustrated in Fig.~\ref{fig:sfe} 
is most likely a consequence of the fact that high-$z$ bins are mostly
populated by galaxies with high SFR, which are characterized by high
SFE, as a consequence of the super-linear slope of the S-K relation.

\begin{figure*}[!t]    
  \resizebox{\hsize}{!}{\includegraphics[angle=90]{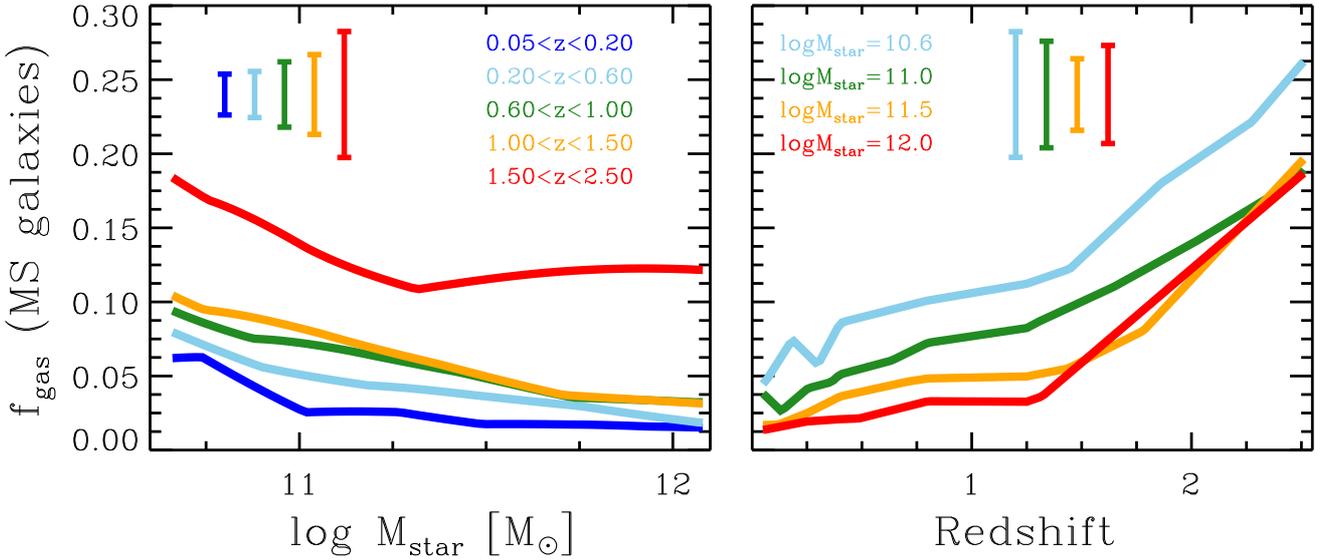}}
\caption{{\it Left}: Gas fraction vs \ms~at different redshifts (in
  different colours according to the legend) for Main Sequence (MS)
  galaxies. {\it Right}: Gas fraction vs redshift at different \ms~(in
  different colours according to the legend) for MS galaxies. Curves
  are obtained by interpolating the \fgas~parameterizations reported
  in Fig.~\ref{fig:fgas_fit} and Table \ref{tab:param} and the MS
  relations (see text for details) at \ms~above the minimum sampled
  \ms~common to all SFR bins. Mean uncertainties on gas fraction
    associated to main sequence galaxies in each redshift ({\it left}) or
    stellar mass ({\it right}) bin 
are plotted.
  }
\label{fig:fgas_ms}
\end{figure*}

We note that the fundamental relation presented here is indeed a
physical result, rather than just a way of looking at the redshift
evolution through the evolution of another parameter (e.g., SFR). 
  In other words, the inclusion of the SFR or stellar mass as
  parameters is not masking a true underlying redshift evolution. As
a matter of fact, no similar relation is obtained if redshift is
replaced to either SFR or \ms.

\subsection{The evolution of the gas fraction among Main Sequence galaxies} \label{sec:fgasms}

The finding that the {\it fundamental $f_{gas}$--$M_{star}$--SFR
  relation} does not evolve with redshift does not contradict the
claimed redshift evolution of the gas fraction in galaxies
\citep[e.g., ][]{daddi10,tacconi10,tacconi13,magdis12}.  Indeed, as
already mentioned, galaxies do not uniformly populate this 3D
surface. As they evolve, the bulk of star-forming galaxies populate
different regions of this surface, as a consequence of gas accretion,
gas consumption by star formation and gas ejection.  The projection of
such a distribution onto the \ms--SFR plane yields the MS and its
evolution with redshift.

As suggested by various models, the evolution of galaxies is likely
driven by the evolution of their gas content. The evolution of the MS
is likely a by-product of the gas content through the S-K relation, or
more generally through the {\it fundamental $f_{gas}$--$M_{star}$--SFR
  relation} illustrated above. While the evolution of the MS has been
constrained by several observations, its driving process, which is the
evolution of the gas content, is still loosely constrained. We can
however exploit the observed evolution of the MS to infer the
evolution of the gas fraction of the population of galaxies dominating
star formation at any epoch, by exploiting the {\it fundamental
  $f_{gas}$--$M_{star}$--SFR relation}.

We take advantage of the mathematical representation of the gas
fraction
 as a function of \ms~at given SFR shown in
Fig.~\ref{fig:fgas_fit}, and we linearly interpolate these relations
onto a finer SFR grid. We then adopt the MS relations reported in
Fig.~\ref{fig:ms3d} and linearly interpolate them onto a fine redshift
grid. At a given \ms~and redshift, we use the MS relation to compute
the expected SFR,
 according to which we 
 select the appropriate
\fgas~parameterization.

The resulting evolution of
 \fgas~with stellar mass at different
redshifts (colour coded) is shown in the left panel of
Fig.~\ref{fig:fgas_ms}. The orthogonal plot, i.e., the redshift
evolution of \fgas~for different stellar masses (colour coded), is
shown in the right panel of Fig.~\ref{fig:fgas_ms}.  These plots
illustrate how the ``bulk'' of the star-forming galaxy population at
various epochs populates the 3D {\it fundamental
  $f_{gas}$--$M_{star}$--SFR relation} as a function of redshift.
Essentially, for a given stellar mass, the average gas content of
star-forming galaxies increases steadily with redshift, at least out
to $z\sim2.5$. The increase rate is steeper for low mass galaxies with
respect to massive galaxies. Galaxies with \msbin~$\sim 10.6$ reach
\fgas~$\sim0.25$ around the peak of cosmic star formation at $z\sim
2.5$, while massive galaxies, with \msbin~$\sim 12$ reach a gas
fraction of only 0.15 at the same cosmic epoch.  This behaviour is
consistent with a downsizing scenario \citep{cowie96,fontanot09},
where massive galaxies have already consumed most of their gas at high
redshift, while less massive galaxies have a larger fraction of gas
(more complex scenarios resulting from the interplay of inflows,
outflows and star formation are not excluded). Further, in massive
galaxies the gas fraction decreases more steeply, moving towards lower
redshift (with respect to low mass galaxies) and their gas evolution
flattens to low values at $z\lesssim 1.3$. Instead, low mass galaxies
show a shallower and more regular decrease of the gas content, moving
towards lower redshifts. Both trends are further indications of
downsizing.

The \fgas~values are somewhat lower by a factor of $\sim$1.5--2 
on average (after accounting for the IMF conversion) than inferred by
the high-$z$ CO survey  of \cite{tacconi13}.  A similar or even larger mismatch
  with CO-based results 
was found by \cite{conselice13}, who compute gas fractions from SFR
and galaxy sizes by
inverting the S-K law. We ascribe the
discrepancy to the combination of the various uncertainties associated
 with CO studies and with our method. 
 In addition, the underestimate by $\simeq$~50\% of the dust mass
  of unresolved 
  sources found by \cite{galliano11} may also explain the lower values
 found by us. 
 The gas fractions derived by us are also lower by a factor of $\sim
 2$ than those published by \cite{magdis12}, who adopt a similar
 method.  This might be caused by cosmic variance effects: based on
 the two GOODS fields only, the analysis of \cite{magdis12} may be
 affected by statistical uncertainty. The inclusion of COSMOS data
 provides much improved statistics that is crucial in stacking
 analyses. Indeed, the stacking result is closely related to the
 number of stacked sources. Even if COSMOS is shallower than the deep
 GOODS fields, SPIRE observations, on which dust masses mostly rely,
 are confusion limited. Therefore, the statistics is strongly
 dominated by COSMOS.  To verify whether cosmic variance effects
   could be responsible for such disagreement, we repeated our
   analysis by only including the two GOODS fields. Given the limited
   statistics, we end up with only 10 data points. We compared these
   with our gas fractions and found that  in 30\% of the cases  the former
  are indeed larger by a factor of 2--2.5, while the rest of the points
    are consistent within their error bars.   Finally, we note
that the disagreement with previous works is reduced when the GRASIL
model is adopted instead of \cite{draineli07}.

\subsection{Comparison with theoretical predictions}\label{sec:models}

The evolution of the gas fraction is a powerful observable to test the
various physical processes at play in galaxies and implemented by
theoretical models, such as star formation, gas cooling and feedback.
Here we compare our findings for the evolution of the gas fraction
with the expectations of the semi-analytical model of galaxy formation
developed by \cite{menci08} (and references therein).  This connects,
within a cosmological framework, the baryonic processes (gas cooling,
star formation, supernova feedback) to the merging histories of the
dark matter haloes, computed by means of a Monte Carlo simulation.
Gas is converted into stars through two main channels: a steady (or
quiescent) accretion mode, in which the cold gas in the galaxy disk is
converted into stars on long timescales ($\sim$1~Gyr), and an
interaction-driven mode, where gas destabilized during major and minor
mergers and fly-by events is converted into stars on shorter
timescales ($\sim$$10^7$~yr; see \citealt{lamastra13,lamastra13b} for
a more detailed description).  AGN activity triggered by the same
galaxy interactions and the related feedback processes are also
included.

\begin{figure}[!t]

  \resizebox{\hsize}{!}{\includegraphics[angle=0]{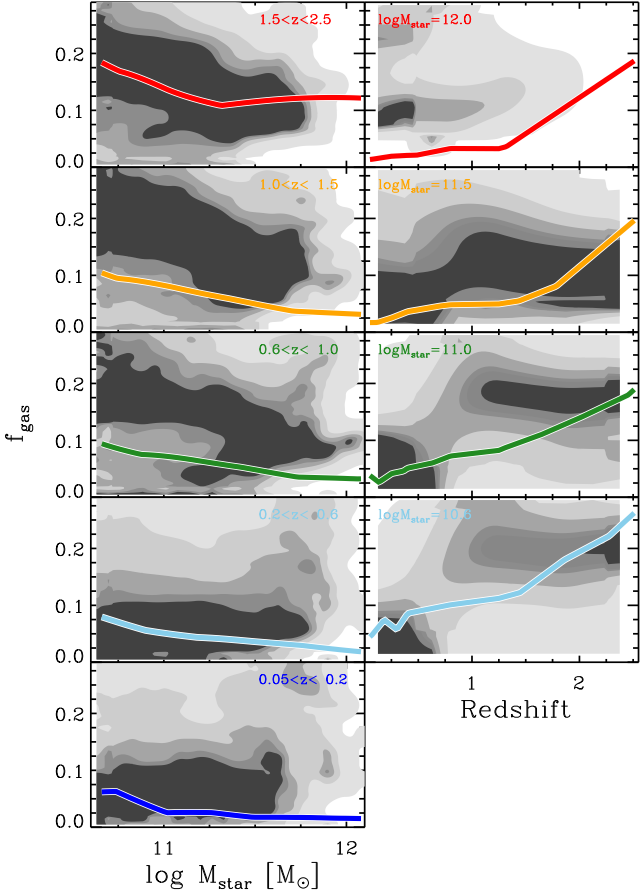}}
  \caption{ Predicted evolution of the gas fraction according to the
    semi-analytical model of \cite{menci08}. The five filled contours
    indicate the fraction of galaxies having a given \fgas~at a fixed
    \ms~({\it left} panels) and redshift ({\it right} panels).  The
    trends for MS galaxies extrapolated by our observations (shown in
    Fig.~\ref{fig:fgas_ms}) are overplotted. }
\label{fig:models}
\end{figure}

The predicted gas fraction as a function of stellar mass and redshift
is shown in Fig.~\ref{fig:models}.  On the same figure we report the
extrapolations for MS galaxies based on our observations already shown
in Fig.~\ref{fig:fgas_ms}. As discussed above, MS galaxies represent
the bulk of the galaxy population and can be directly compared to the
darkest contours, enclosing the region occupied by most of the
galaxies.

Observations are generally well reproduced by the theoretical model,
although with some systematic deviations.  The trends with both
stellar mass and redshift are recovered, as well as the downsizing
expectations: a strong evolution can be noticed in low mass galaxies
(\ms $\lesssim 10^{11} \rm M_\odot/yr$), which are gas-rich out to
$z\sim 1$ (bottom right panel), while progressively more massive
galaxies have already consumed their gas at this epoch (upper right
panel).  While a very good agreement is recovered for all stellar
masses at high redshift ($z\sim 2$, upper-left panel), the predicted
evolution of the gas fraction is more regular than observed at
intermediate redshifts, with a gas fraction in \msbin $\lesssim 11.5$
galaxies of $\sim 0.2$ at $z\gtrsim 0.6$, around twice the observed
value (central left panels).  The overall systematic gas richness of
model galaxies compared to the observations relates to the
long-standing problem of theoretical models in reproducing the galaxy
stellar mass functions at high redshift. Indeed, the number of massive
galaxies is underpreticted by the models
\citep[e.g.][]{fontanot09,santini12a}, consistently with the
inefficiency of the gas conversion and mass buildup processes in the
distant Universe. Once gas consumption has started, it is not
efficiently suppressed at late stages. Indeed, the model predicts a
fraction of very massive (\msbin$\gtrsim 11.5$) galaxies which are
still gas-rich at $z<1$, at variance with what observed (lower- and
central-left panels and top-right one). Although it can be partly
ascribed to fluctuations in the \fgas~distribution generated by the
low number statistics of such high \ms~galaxies, this behaviour is a
manifestation of a known problem common to all theoretical models, in
which the suppression of the star formation activity is still
inefficient, despite the feedback processes at work. This is related
to the difficulties in reproducing the fraction of red passive
galaxies \citep{fontana09}.

For all these reasons, the comparison of observed and modeled gas
fraction is of major importance to constrain the physical processes
implemented in models of galaxy formation and evolution.  A more
detailed and complete comparison with theoretical expectations will be
tackled in a future work.

\section{Summary}\label{sec:summ}

We have used Herschel data from both PACS and SPIRE imaging cameras
to
 estimate the dust mass of a large sample of galaxies extracted
from the
 GOODS-S, GOODS-N and COSMOS fields.  To explore a wide
range of galaxy properties, including low mass and moderate
star-forming galaxies, we have performed a stacking analysis on a grid
of redshifts, stellar
 masses and SFR, and considered average values.
With these outputs we have studied the scaling relations in place
between the dust content of galaxies and their stellar mass and SFR at
different redshifts, from the local Universe out to
 $z=2.5$.  Our
main results are the following.

\begin{itemize}
\item[$\bullet$] No clear evolution of the dust mass with redshift is
  observed at a given SFR $and$ stellar mass. Although there is a global
  redshift evolution of the dust content in galaxies, as a consequence
  of the increased ISM content at high-$z$, our findings indicate that
  galaxies with the same properties (same SFR $and$ same \ms) do not
  show any significant difference in terms of dust content across the
  cosmic epochs, at least out to $z\sim2.5$. In other words dust mass
  in galaxies is mostly determined by SFR and \ms~and is
  independent of redshift.
\item[$\bullet$] The dust content is tightly correlated with the star
  formation
 activity of the galaxy. This correlation is in place at
  all values of \ms~probed and at least out to $z \sim 2.5$. Under the
  assumption that
 the dust content is proportional to the gas
  content (with a factor
 scaling with the gas metallicity), the
  observed correlation is a
 natural consequence of the
  Schmidt-Kennicutt (S-K) law.
\item[$\bullet$] The correlation between the dust and stellar mass
  observed by previous studies (which averaged together all SFR)
  becomes much flatter or even disappears when taken at a fixed
  SFR. The $\rm M_{dust}-M_{star}$ relation is at least partly a
  result of the \md--SFR
 correlation combined with the Main Sequence
  (MS) of star-forming
 galaxies. 
\end{itemize}

We have then taken one step  further 
and computed gas metallicities from the
 stellar mass and the SFR
according to the Fundamental Metallicity
 Relation (FMR) fitted by
\cite{mannucci10}, and estimated gas masses by assuming that
 the
dust-to-gas ratio linearly scales with the gas metallicity. We note
that all our results are robust against the specific parameterization
chosen to describe the gas metallicity (e.g., FMR against
redshift-dependent mass-metallicity relation). This method
 provides
a complementary approach to investigate the galaxy gas
 content
independently of CO observations.
Under our
 assumptions we find the following.

\begin{itemize}
\item[$\bullet$] We fit a power law relation between the SFR and the
  gas mass, in good 
  agreement 
 with that previously obtained by \cite{daddi10}, and
  also broadly
 consistent with the results of \cite{genzel10}. This
  agreement is
 remarkable, given the 
 completely different
  approach between our study and the two works
 above based on CO
  measurements. We find that all galaxies follow the same star
  formation law (integrated S-K law), with no evidence of starbursts
  lying on an offset relation, though our sample lacks the most
  extremely starbursting sources (such as local ULIRGs and their
    analogues at high-$z$). The slope of this relation is on average
  steeper than unity, implying that strongly star-forming galaxies
  have higher star formation efficiency (SFE, i.e., the inverse
    of the depletion time), or shorter depletion
    time.  We also find a mild, but significant evolution
  of the S-K law with redshift.
\item[$\bullet$] We observe an evolution of
  the 
  SFE with redshift, by about a factor of 10 from $z\sim0$ to
  $z\sim2.5$. This applies to the bulk of the galaxy population
  dominating star formation at each epoch.  However, it is not clear
  whether such evolution is an intrinsic redshift evolution or is
  simply a consequence of sampling more star-forming galaxies at high
  redshift combined with the slope of the integrated S-K relation
  being on average steeper than unity.
\item[$\bullet$] The measured gas fraction decreases with stellar mass
  and
 increases with SFR, as expected. However, when considering
  bins of
 given stellar mass {\it and} SFR, the gas fraction does
  not show any
 redshift evolution, at least out to $z\sim2.5$.  This
  primarily results from the non-evolution of the dust mass (within uncertainties), with gas
  metallicity effects only providing a second-order contribution.
  The 3D relation between \fgas, \ms~and SFR is a {\it fundamental
    relation} that holds at any redshift.  It provides a powerful tool
  to overcome the observational difficulties related with the
  measurement of gas or dust masses and to analyse the gas content for
  much larger samples of galaxies.  Galaxies populate such a 3D {\it
    fundamental $f_{gas}$--$M_{star}$--SFR relation} in a different
  way throughout the cosmic epochs.  The distribution of galaxies on
  the 3D {\it fundamental relation} onto the \ms--SFR plane gives the
  MS and its evolution with redshift.
\item[$\bullet$] We ``de-project'' the MS galaxies, at various cosmic
  epochs, onto the 3D {\it fundamental $f_{gas}$--$M_{star}$--SFR
    relation}, to infer the evolution of the gas fraction of
  ``typical'' star-forming galaxies as a function of redshift.  A
  clear redshift evolution from $z\sim0$ to $z\sim 2.5$ in
 the gas
  fraction is observed for MS galaxies.  The evolution of the gas
  content in massive (\ms~$\gtrsim$~10$^{11} \rm M_{\odot}$) galaxies
  is steep between $z\sim2.5$ and $z\sim1.2$ and flattens to low
  \fgas~ values at lower redshifts. Low mass \ms $\lesssim$~10$^{11} \rm M_{\odot}$ galaxies show a less steep and more regular decrease
  of the gas fraction from $z\sim2.5$ to $z\sim 0$. These trends are
  in agreement with the downsizing scenario for galaxy evolution, and
  they are on average well reproduced by the theoretical
  expectations of the semi-analytical model of \cite{menci08}, 
    despite a systematic larger gas richness compared to our
    data.
\end{itemize}

\begin{acknowledgements}
PS thanks N. Scoville for interesting and useful discussions and 
A. Marconi and G. Risaliti for helping with fitting
routines and statistical issues.  This work was supported by grant ASI
I/005/11/0.  PACS has been developed by a consortium of institutes led
by MPE (Germany) and including UVIE (Austria); KU Leuven, CSL, IMEC
(Belgium); CEA, LAM (France); MPIA (Germany); INAF-IFSI/ OAA/OAP/OAT,
LENS, SISSA (Italy); IAC (Spain). This development has been supported
by the funding agencies BMVIT (Austria), ESA-PRODEX (Belgium),
CEA/CNES (France), DLR (Germany), ASI/INAF (Italy), and CICYT/MCYT
(Spain).  SPIRE has been developed by a consortium of institutes led
by Cardiff University (UK) and including University of Lethbridge
(Canada), NAOC (China), CEA, LAM (France), IFSI, University of Padua
(Italy), IAC (Spain), Stockholm Observatory (Sweden), Imperial College
London, RAL, UCL-MSSL, UKATC, University of Sussex (UK), Caltech, JPL,
NHSC, University of Colorado (USA). This development has been
supported by national funding agencies: CSA (Canada); NAOC (China);
CEA, CNES, CNRS (France); ASI (Italy); MCINN (Spain); SNSB (Sweden);
STFC, UKSA (UK); and NASA (USA).
\end{acknowledgements}
\bibliographystyle{aa}

\begin{appendix}

\section{Statistics on the $z$--\ms--SFR grid} \label{app:statistics}

We report in Tables \ref{tab:stat1} to \ref{tab:stat5} the number of
sources 
in each $z$--\ms--SFR bin and the  associated average dust mass.

\begin{table*}
\centering
\caption{Number of sources (upper number in each cell) and average
  dust mass (lower number in each cell) in each $z$--\ms--SFR bin.} 
\input{tab_stat_Mdust_0.txt}
\tablefoot{Masses are in ${\rm M_\odot}$ and SFR are in ${\rm
    M_\odot/yr}$. The three numbers in parentheses in the middle
    row of each table cell show the contribution of GOODS-S, GOODS-N and
    COSMOS fields, respectively, to the bin. The bin with the lowest
    SFR is never populated after all selections applied (see Sect.~\ref{sec:mdust}).}
\label{tab:stat1}
\end{table*}
\begin{table*}
\centering
\caption{Same as Table \ref{tab:stat1}.}
\input{tab_stat_Mdust_1.txt}
 \label{tab:stat2}
\end{table*}
\begin{table*}
\centering
\caption{Same as Table \ref{tab:stat1}.}
\input{tab_stat_Mdust_2.txt}
 \label{tab:stat3}
\end{table*}
\begin{table*}
\centering
\caption{Same as Table \ref{tab:stat1}.}
\input{tab_stat_Mdust_3.txt}
 \label{tab:stat4}
\end{table*}
\begin{table*}
\centering
\caption{Same as Table \ref{tab:stat1}.}
\input{tab_stat_Mdust_4.txt}
 \label{tab:stat5}
\end{table*}

\section{Reliability of the SFR estimates}\label{app:cfrsfr}

\begin{figure*}[!t]
\resizebox{\hsize}{!}{
\includegraphics[angle=0]{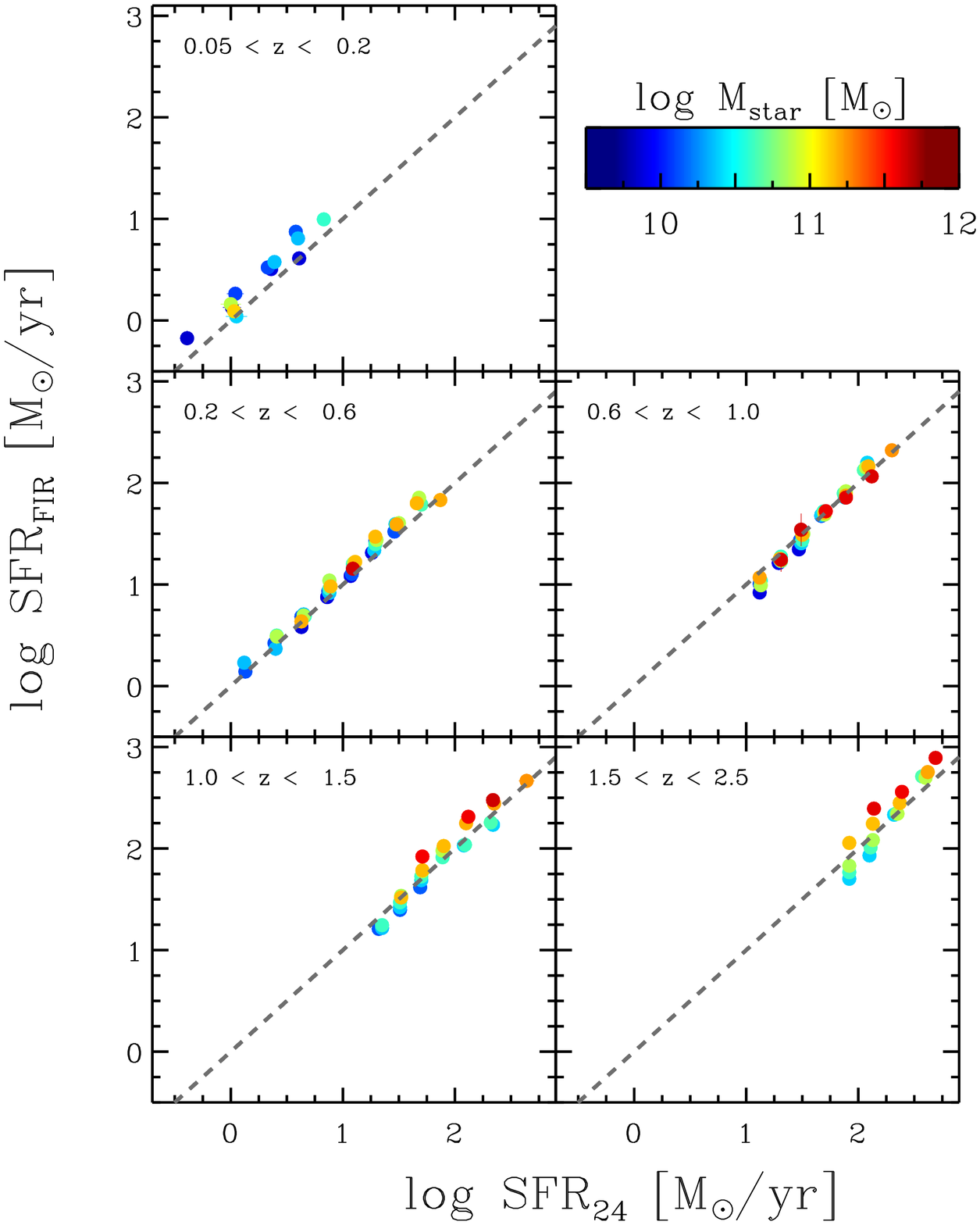}
\includegraphics[angle=0]{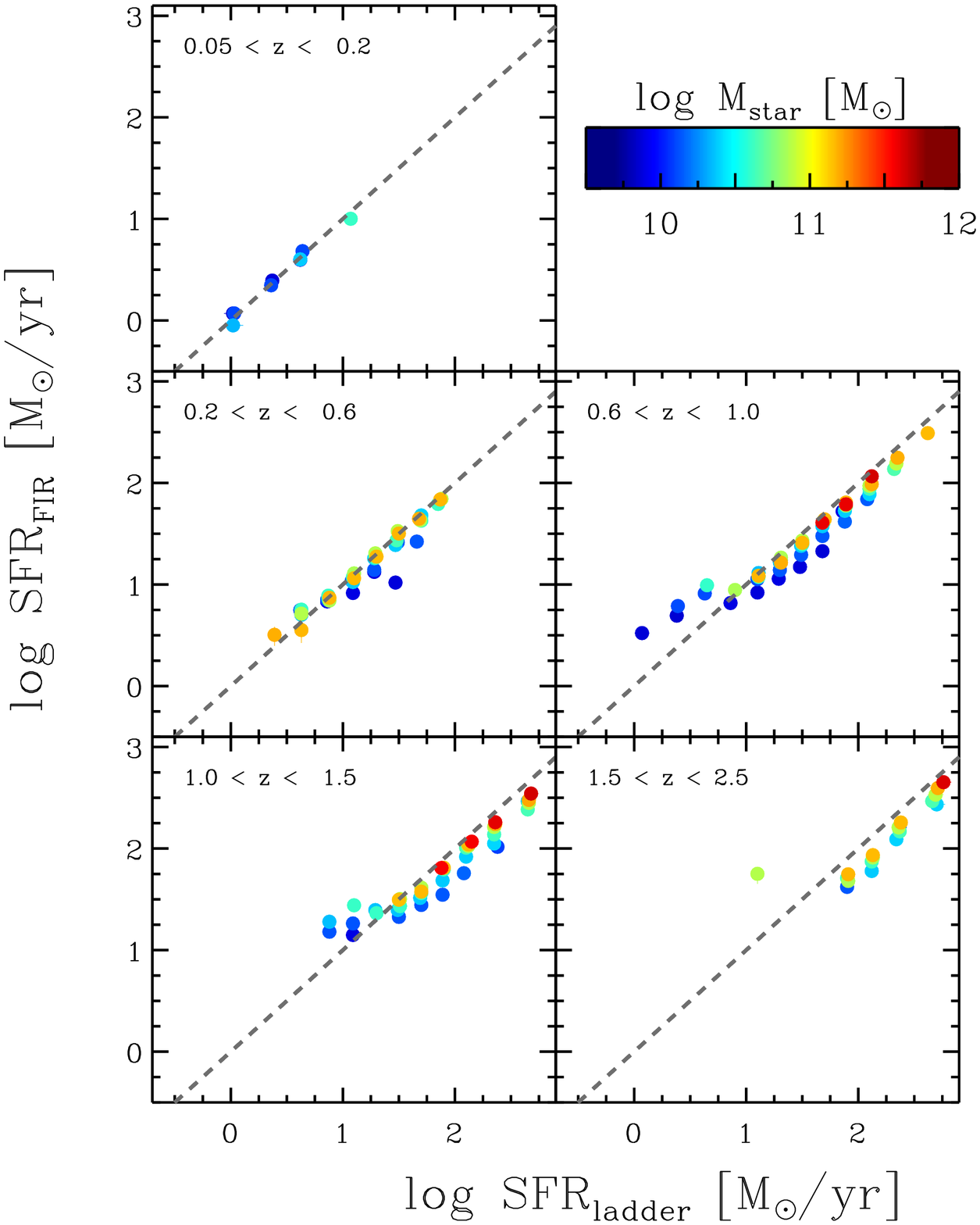}
}
\caption{{\it Left}: Comparison between the  24~\mic-based SFR used
    in the analysis ($x$-axis)
 and the SFR measured 
 by  fitting the average FIR stacked flux densities for each bin of the grid
  ($y$-axis). The
  colour code indicates the mean stellar mass in each bin. {\it Right}:
   Same as the left panel, but FIR-based SFR are compared to the SFR
  measured 
  by means of a ``ladder" approach (see text). }
\label{fig:cfrsfr}
\end{figure*}

To 
verify the reliability of the 24~\mic-based SFR
tracer (SFR$_{24}$), 
we compare it with the SFR measured by fitting
the average FIR stacked flux densities ($\rm SFR_{FIR}$). We fit these
flux densities
with the \cite{dh02} IR template library  to infer an estimate of
  the total IR luminosity, and account for the unobscured SFR by taking into account the
  average rest-frame UV luminosity in the bin uncorrected for
  extinction (see Sect. \ref{sec:sfr} for 
more details). The same results are found if the MS template of
  \cite{elbaz11} is used instead of the \cite{dh02} library. In the left panel of Fig.~\ref{fig:cfrsfr} we compare
the average  SFR$_{24}$ in each bin of the grid with the average $\rm SFR_{FIR}$.  
The two SFR measurements nicely agree with each other
with the only noticeable exception of the lowest redshift bin, making us
confident of the method adopted. The small offset observed at
  low-$z$ does not significantly affect our results. 

  We also repeated the same test by making use of a ``ladder of SFR
  indicators'' \citep[$\rm SFR_{ladder}$, e.g.][]{wuyts11a} as input
  for the grid production. Such a ``ladder'' approach combines
  different SFR estimates by using the best available choice for each
  galaxy. More specifically, a Herschel-based SFR is used for galaxies
  detected by PACS or SPIRE, the 24~\mic-based tracer is adopted for
  galaxies undetected by Herschel but detected by MIPS, and the output
  of the optical-UV SED fitting described in Sect. \ref{sec:mstar} is
  used for galaxies undetected at IR wavelengths. Most importantly,
  this approach has the advantage of increasing the number of galaxies
  for which a SFR estimate is available and enlarging the SFR
  dynamical range. However, as evident from the right panel of
  Fig.~\ref{fig:cfrsfr}, the scatter with respect to $\rm SFR_{FIR}$
  is larger than in the previous case. Moreover, the correlation
  between $\rm SFR_{ladder}$ and $\rm SFR_{FIR}$ flattens
at low SFR and $z>0.2$, exactly at the SFR regime where in principle
  the ``ladder'' approach provides an improvement over the
  24~\mic-based SFR.  One possibility to explain the flat behaviour at
  low SFR (below a redshift-dependent threshold) shown in the right
  panel of Fig.~\ref{fig:cfrsfr} is to ascribe it to failures in the
  associations of optical counterparts for the extremely faint IR
  galaxies or blending issues mostly affecting the faintest galaxies
  during the stacking procedure. Alternatively, dust heating by old
  stellar population might also be responsible for the enhanced IR
  flux at low SFR. However, investigating the reasons of such
  disagreement is beyond the scope of the present work.  Based on the
  tests performed, we decide to use SFR$_{24}$ as a SFR tracer, at
  the expenses of reducing the SFR dynamical range, in order not to
  run the risk to introduce systematics in the analysis.

\section{Simulation to test against possible degeneracies in the SFR--\md~correlation}\label{app:simul}

We run a simulation to verify that the trend observed between the
  SFR and the dust mass is real and not a trivial outcome of the fact
  that both physical variables are related to the FIR peak of the
  galaxy SED. Indeed, while the SFR is simply proportional to the
  integrated light in the dust emission peak, the dust mass depends
  not only on the normalization of the spectrum but also on the
  temperature of the grains, which determines its shape.

\begin{figure*}[!t]
\centering
\resizebox{0.8\hsize}{!}{\includegraphics[angle=90]{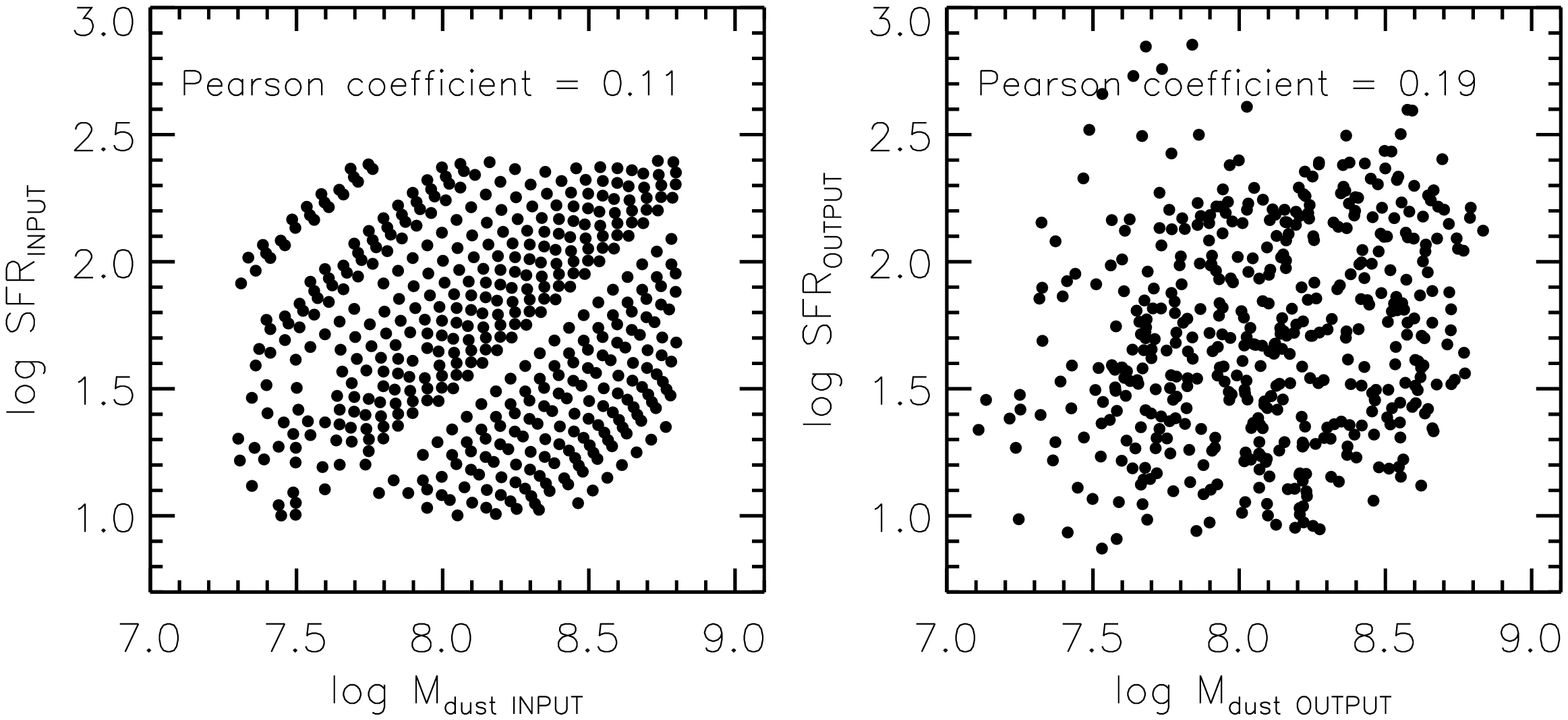}}
\caption{Distribution of mock SFR and \md~in input ({\it left} panel)
  and output ({\it right} panel) of the simulation described in
  text. The measure of the SFR and \md~does not introduce a
 correlation in an initially uncorrelated distribution. The value of
 the Pearson coefficient is printed in both cases.}
\label{fig:simul}
\end{figure*}

The aim of the simulation is to verify whether an initially scattered
and uncorrelated distribution of mock SFR and \md~gives rise to a
correlation when the two quantities are recomputed according to our
methods. To do that, we consider a set of GRASIL templates, each
associated to a dust mass ($\rm M_{dust\ INPUT}$) and to a SFR ($\rm
SFR_{ INPUT}$). The latter is computed by integrating the template
from 8 and 1000~\mic~and multiplying by the calibration factor
$1.8\times 10^{-10}$ (see Sect. \ref{sec:sfr}). In order to sample a
wide region of the SFR--\md~parameter space, we multiply each
SFR--\md~pair by a normalization factor. This corresponds to
multiplying the associated SED, since both parameters scale with the
SED normalization. We consider a range in SFR and \md~which mimic that
observed in one of the redshift interval most populated by our data,
i.e., $0.6<z<1$.  The resulting SFR--\md~distribution is shown in the
left panel of Fig.~\ref{fig:simul}.

We redshift each mock galaxy to a random redshift within the 0.6-1
interval, and interpolate the associated spectrum with the MIPS
24~\mic~and Herschel 100--500~\mic~filters. To mimic the real case, we
perturb such mock flux densities by adding a noise randomly extracted
from the observed noise distribution in each band. We then measure the
SFR ($\rm SFR_{ OUTPUT}$) and \md~($\rm M_{dust\ OUTPUT}$) for each
mock galaxy exactly as we have done for the real data. Consistently
with what described in Sect. \ref{sec:mdust}, we reject sources not
compliant with our requirements to ensure reliable dust mass
estimates.  The resulting measurements are shown in the right panel of
Fig.~\ref{fig:simul} and show no evidence for any correlation between
SFR and \md. The absence of any correlation induced by our measures is
statistically confirmed by the values of the Pearson coefficients on
the input (0.11) and output (0.19) data point distributions.  This
simulation illustrates that the correlations between \md~and SFR is
not an artefact of the method, but is real (i.e., the result of the
S-K law).

\section{Fits of the far-IR SEDs}\label{app:mdustfits}

\begin{figure*}[!t]
\resizebox{\hsize}{!}{
\includegraphics[angle=0]{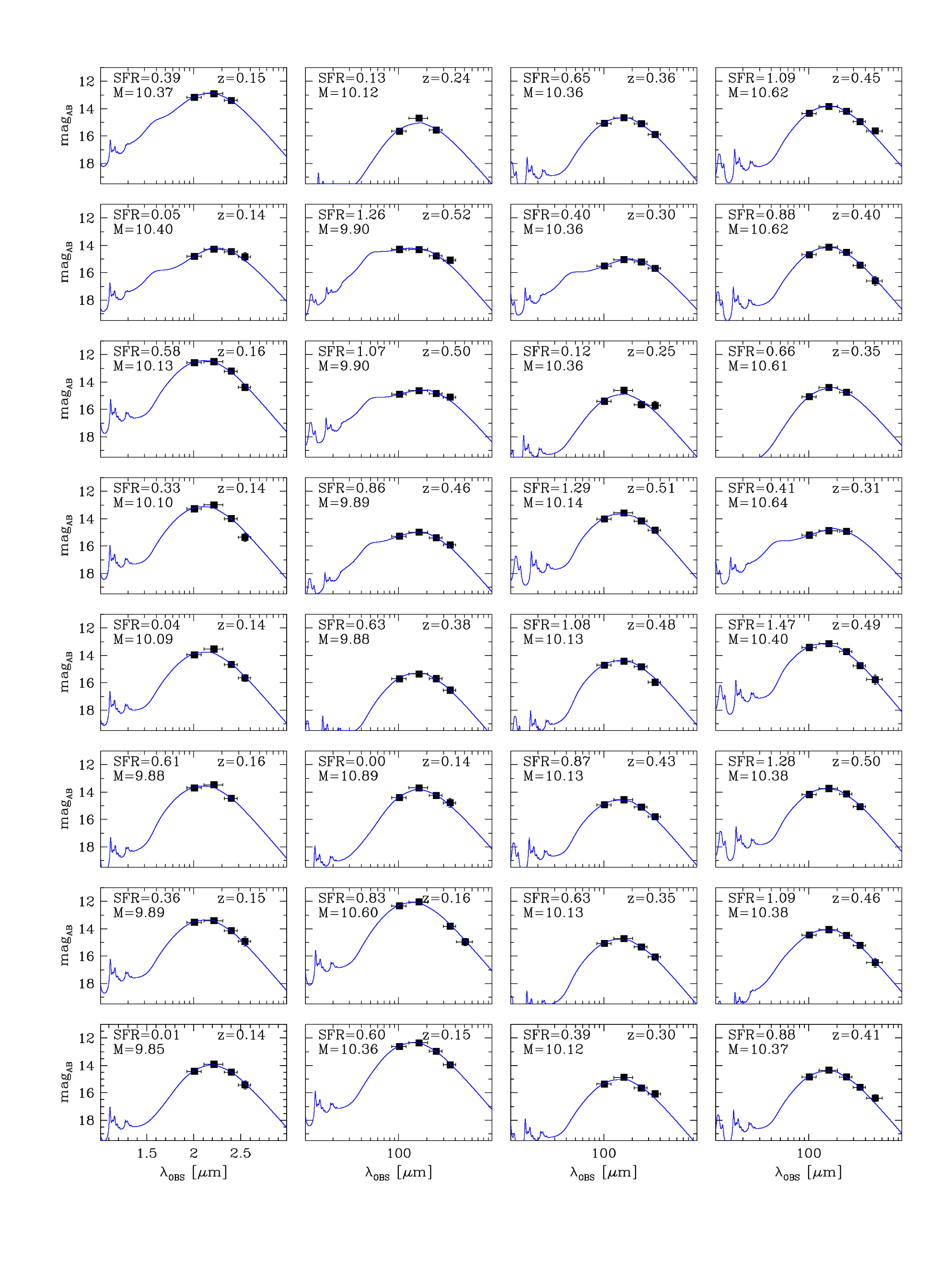}
}
\caption{Far-IR SED fits of the Herschel stacked flux densities with
  \cite{draineli07} templates. Each panel refers to a bin of the $z$-\ms-SFR
  grid. The average value of redshift, stellar
  mass and SFR for galaxies belonging to each bin is printed in each panel. }
\label{fig:fits1}
\end{figure*}

\begin{figure*}[!t]
\resizebox{\hsize}{!}{
\includegraphics[angle=0]{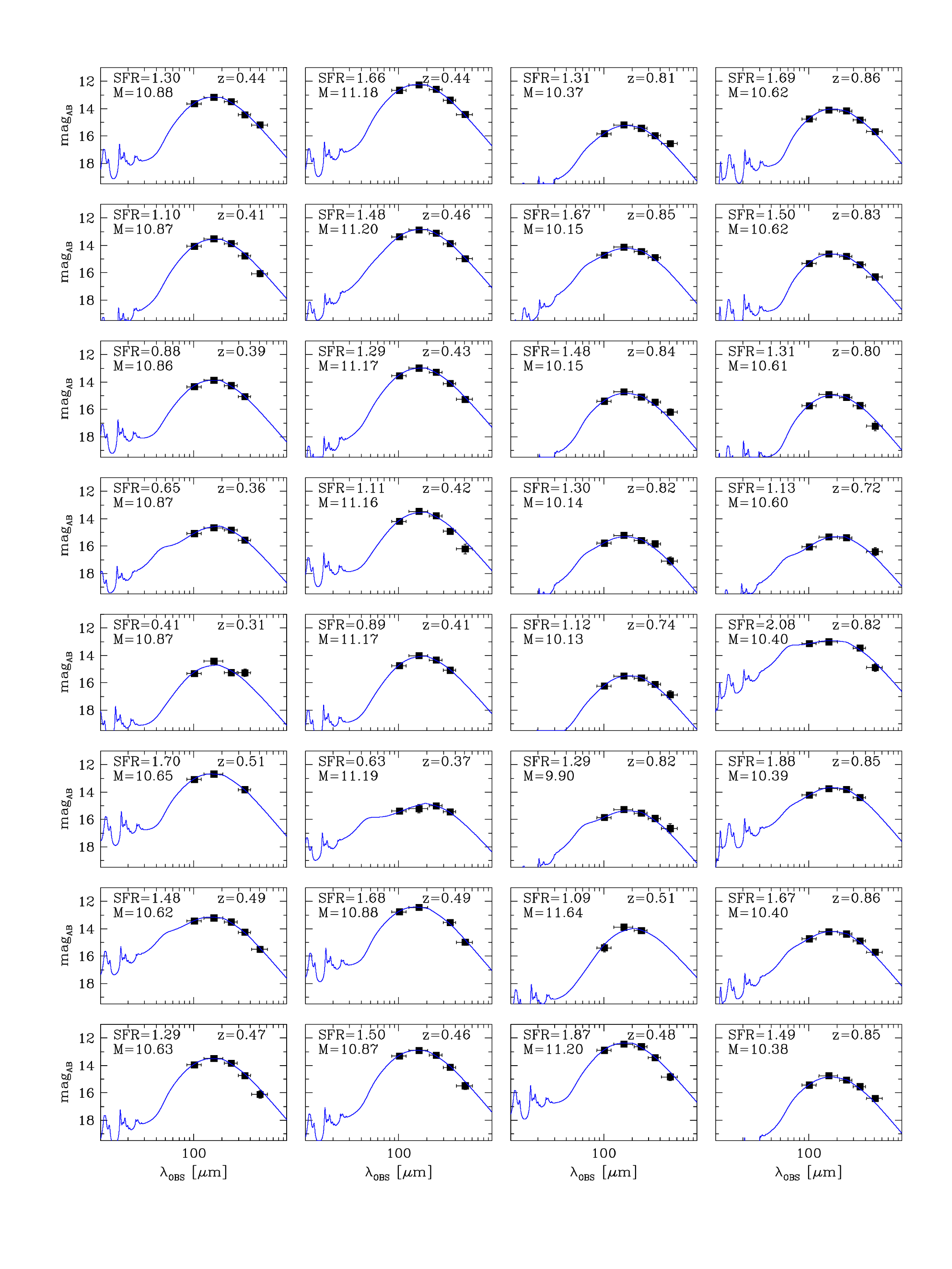}
}
\caption{Same as Fig.~\ref{fig:fits1} (continued). }
\label{fig:fits2}
\end{figure*}

\begin{figure*}[!t]
\resizebox{\hsize}{!}{
\includegraphics[angle=0]{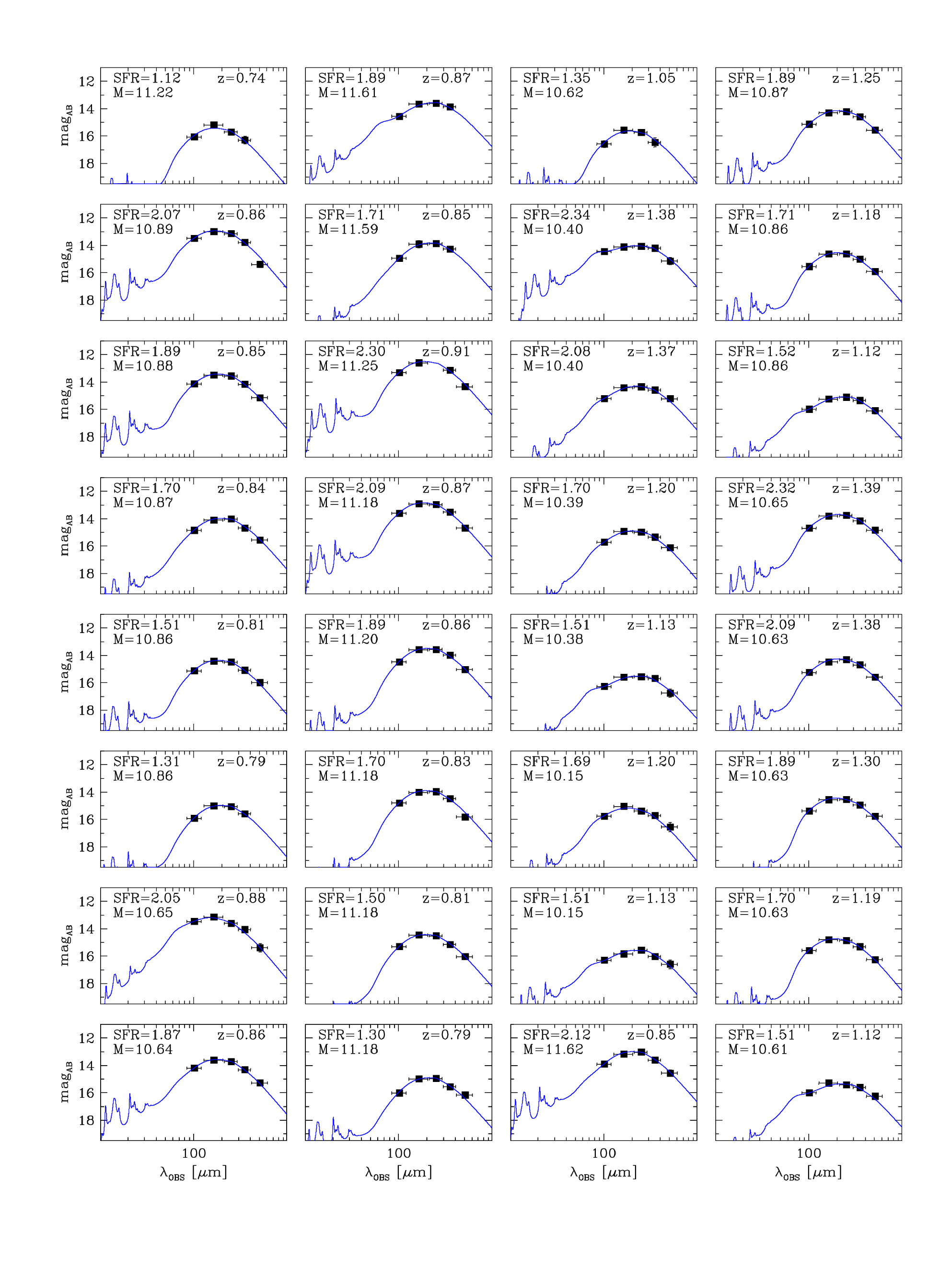}
}
\caption{Same as Fig.~\ref{fig:fits1} (continued).  }
\label{fig:fits3}
\end{figure*}

\begin{figure*}[!t]
\resizebox{\hsize}{!}{
\includegraphics[angle=0]{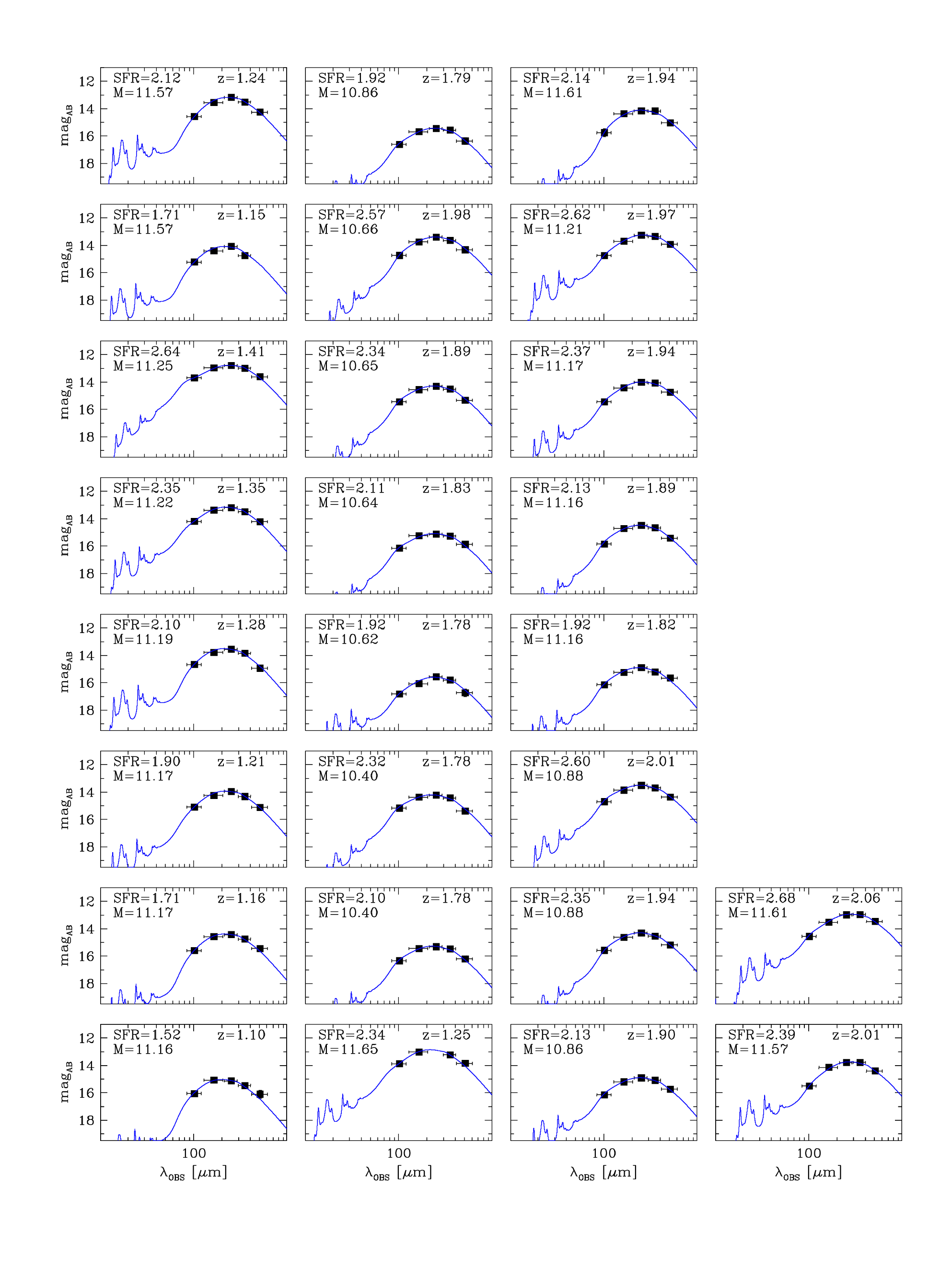}
}
\caption{Same as Fig.~\ref{fig:fits1} (continued).  }
\label{fig:fits4}
\end{figure*}

In Figs.~\ref{fig:fits1} to \ref{fig:fits4} we report the best fits of
Herschel stacked flux densities with \cite{draineli07} templates
computed to estimate the dust mass.   The secondary bump around
  50~\mic~which can be seen in few of the best-fit SEDs is due to a
  warm dust component. This feature gradually disappears when the
  maximum radiation intensity ($U_{max}$) in the \cite{draineli07}
  model is set to lower values. However, fixing $U_{max}$ to a value
  lower than $10^6$ has the overall effect of making each template
  slightly warmer. This has the effect of increasing the inferred dust
  masses by a factor of 1.5--2, due to a larger normalization of the
  SED for a given set of observed fluxes. We decided to follow the
  prescription given by \cite{draine07} and fixing $U_{max}$ to
  $10^6$. However, we note that an offset would not change our main
  results.  


\end{appendix}

\end{document}

%% file: tab_stat_Mdust_0.txt
\begin{tabular} {cccccccc}
\hline \noalign{\smallskip} 
\multicolumn{8}{c}{$  0.05\leq z<  0.20$}\\
\noalign{\smallskip}\hline\noalign{\smallskip}
$\rm \log SFR$ & \multicolumn{7}{c}{$\rm \log M_{star}$}\\
\noalign{\smallskip}\noalign{\smallskip}
 & 9.75 -- 10.00
 & 10.00 -- 10.25
 & 10.25 -- 10.50
 & 10.50 -- 10.75
 & 10.75 -- 11.00
 & 11.00 -- 11.50
 & 11.50 -- 12.00
 \\ 
\noalign{\smallskip}\hline\noalign{\smallskip}
\noalign{\smallskip}
$-0.75$ -- $-0.25$
 &  
 &  
 &  
 &  
 &  
 &  
 &  
\\
 &  
 &  
 &  
 &  
 &  
 &  
 &  
\\
 &  
 &  
 &  
 &  
 &  
 &  
 &  
\\
\noalign{\smallskip}
$-0.25$ -- $0.25$
 &     36
 &     33
 &     16
 &  
 &    193
 &  
 &  
\\
 & (     0,     1,    35)
 & (     0,     0,    33)
 & (     1,     0,    15)
 &  
 & (    14,    19,   160)
 &  
 &  
\\
 &     7.11$_{-0.15}^{+0.26}$
 &     6.83$_{-0.13}^{+0.13}$
 &     7.42$_{-0.43}^{+0.07}$
 &  
 &     7.31$_{-0.23}^{+0.26}$
 &  
 &  
\\
\noalign{\smallskip}
$0.25$ -- $0.50$
 &     12
 &     10
 &     16
 &  
 &  
 &  
 &  
\\
 & (     0,     0,    12)
 & (     0,     0,    10)
 & (     0,     1,    15)
 &  
 &  
 &  
 &  
\\
 &     7.06$_{-0.13}^{+0.31}$
 &     7.05$_{-0.08}^{+0.12}$
 &     7.55$_{-0.14}^{+0.30}$
 &  
 &  
 &  
 &  
\\
\noalign{\smallskip}
$0.50$ -- $0.75$
 &     22
 &     12
 &     10
 &  
 &  
 &  
 &  
\\
 & (     1,     0,    21)
 & (     0,     1,    11)
 & (     0,     1,     9)
 &  
 &  
 &  
 &  
\\
 &     6.95$_{-0.19}^{+0.30}$
 &     7.44$_{-0.07}^{+0.05}$
 &     7.64$_{-0.06}^{+0.04}$
 &  
 &  
 &  
 &  
\\
\noalign{\smallskip}
$0.75$ -- $1.00$
 &  
 &  
 &  
 &    142
 &  
 &  
 &  
\\
 &  
 &  
 &  
 & (     4,    13,   125)
 &  
 &  
 &  
\\
 &  
 &  
 &  
 &     7.71$_{-0.10}^{+0.05}$
 &  
 &  
 &  
\\
\noalign{\smallskip}
$1.00$ -- $1.20$
 &  
 &  
 &  
 &  
 &  
 &  
 &  
\\
 &  
 &  
 &  
 &  
 &  
 &  
 &  
\\
 &  
 &  
 &  
 &  
 &  
 &  
 &  
\\
\noalign{\smallskip}
$1.20$ -- $1.40$
 &  
 &  
 &  
 &  
 &  
 &  
 &  
\\
 &  
 &  
 &  
 &  
 &  
 &  
 &  
\\
 &  
 &  
 &  
 &  
 &  
 &  
 &  
\\
\noalign{\smallskip}
$1.40$ -- $1.60$
 &  
 &  
 &  
 &  
 &  
 &  
 &  
\\
 &  
 &  
 &  
 &  
 &  
 &  
 &  
\\
 &  
 &  
 &  
 &  
 &  
 &  
 &  
\\
\noalign{\smallskip}
$1.60$ -- $1.80$
 &  
 &  
 &  
 &  
 &  
 &  
 &  
\\
 &  
 &  
 &  
 &  
 &  
 &  
 &  
\\
 &  
 &  
 &  
 &  
 &  
 &  
 &  
\\
\noalign{\smallskip}
$1.80$ -- $2.00$
 &  
 &  
 &  
 &  
 &  
 &  
 &  
\\
 &  
 &  
 &  
 &  
 &  
 &  
 &  
\\
 &  
 &  
 &  
 &  
 &  
 &  
 &  
\\
\noalign{\smallskip}
$2.00$ -- $2.25$
 &  
 &  
 &  
 &  
 &  
 &  
 &  
\\
 &  
 &  
 &  
 &  
 &  
 &  
 &  
\\
 &  
 &  
 &  
 &  
 &  
 &  
 &  
\\
\noalign{\smallskip}
$2.25$ -- $2.50$
 &  
 &  
 &  
 &  
 &  
 &  
 &  
\\
 &  
 &  
 &  
 &  
 &  
 &  
 &  
\\
 &  
 &  
 &  
 &  
 &  
 &  
 &  
\\
\noalign{\smallskip}
$2.50$ -- $3.00$
 &  
 &  
 &  
 &  
 &  
 &  
 &  
\\
 &  
 &  
 &  
 &  
 &  
 &  
 &  
\\
 &  
 &  
 &  
 &  
 &  
 &  
 &  
\\
\noalign{\smallskip}
\noalign{\smallskip}\hline  
\end{tabular}

%% file: tab_stat_Mdust_1.txt
\begin{tabular} {cccccccc}
\hline \noalign{\smallskip} 
\multicolumn{8}{c}{$  0.20\leq z<  0.60$}\\
\noalign{\smallskip}\hline\noalign{\smallskip}
$\rm \log SFR$ & \multicolumn{7}{c}{$\rm \log M_{star}$}\\
\noalign{\smallskip}\noalign{\smallskip}
 & 9.75 -- 10.00
 & 10.00 -- 10.25
 & 10.25 -- 10.50
 & 10.50 -- 10.75
 & 10.75 -- 11.00
 & 11.00 -- 11.50
 & 11.50 -- 12.00
 \\ 
\noalign{\smallskip}\hline\noalign{\smallskip}
\noalign{\smallskip}
$-0.75$ -- $-0.25$
 &  
 &  
 &  
 &  
 &  
 &  
 &  
\\
 &  
 &  
 &  
 &  
 &  
 &  
 &  
\\
 &  
 &  
 &  
 &  
 &  
 &  
 &  
\\
\noalign{\smallskip}
$-0.25$ -- $0.25$
 &  
 &     70
 &     71
 &  
 &  
 &  
 &  
\\
 &  
 & (     1,     1,    68)
 & (     5,     4,    62)
 &  
 &  
 &  
 &  
\\
 &  
 &     7.06$_{-0.22}^{+0.37}$
 &     7.10$_{-0.21}^{+0.24}$
 &  
 &  
 &  
 &  
\\
\noalign{\smallskip}
$0.25$ -- $0.50$
 &  
 &     38
 &     29
 &     35
 &     58
 &  
 &  
\\
 &  
 & (     0,     3,    35)
 & (     1,     3,    25)
 & (     0,     2,    33)
 & (     1,     4,    53)
 &  
 &  
\\
 &  
 &     7.02$_{-0.21}^{+0.38}$
 &     7.69$_{-0.55}^{+0.12}$
 &     7.81$_{-0.64}^{+0.14}$
 &     7.36$_{-0.14}^{+0.29}$
 &  
 &  
\\
\noalign{\smallskip}
$0.50$ -- $0.75$
 &    159
 &     89
 &     31
 &     16
 &     20
 &     21
 &  
\\
 & (     1,     5,   153)
 & (     0,     3,    86)
 & (     2,     1,    28)
 & (     0,     2,    14)
 & (     0,     0,    20)
 & (     1,     0,    20)
 &  
\\
 &     7.10$_{-0.24}^{+0.53}$
 &     7.16$_{-0.16}^{+0.32}$
 &     7.32$_{-0.14}^{+0.43}$
 &     7.61$_{-0.17}^{+0.43}$
 &     7.64$_{-0.29}^{+0.38}$
 &     7.94$_{-0.54}^{+0.11}$
 &  
\\
\noalign{\smallskip}
$0.75$ -- $1.00$
 &    246
 &    179
 &     77
 &     32
 &     44
 &     12
 &  
\\
 & (     3,     8,   235)
 & (     5,     2,   172)
 & (     1,     3,    73)
 & (     0,     3,    29)
 & (     1,     6,    37)
 & (     0,     1,    11)
 &  
\\
 &     7.43$_{-0.35}^{+0.35}$
 &     7.32$_{-0.12}^{+0.37}$
 &     7.49$_{-0.09}^{+0.17}$
 &     7.62$_{-0.07}^{+0.12}$
 &     7.72$_{-0.10}^{+0.19}$
 &     7.84$_{-0.11}^{+0.13}$
 &  
\\
\noalign{\smallskip}
$1.00$ -- $1.20$
 &    287
 &    326
 &    248
 &     62
 &     52
 &     10
 &     43
\\
 & (    10,    16,   261)
 & (     5,    18,   303)
 & (     1,     3,   244)
 & (     1,     1,    60)
 & (     2,     0,    50)
 & (     0,     0,    10)
 & (     0,     1,    42)
\\
 &     7.97$_{-0.50}^{+0.26}$
 &     7.46$_{-0.11}^{+0.22}$
 &     7.66$_{-0.14}^{+0.21}$
 &     7.84$_{-0.07}^{+0.09}$
 &     7.94$_{-0.10}^{+0.11}$
 &     8.03$_{-0.13}^{+0.11}$
 &     8.29$_{-0.29}^{+0.40}$
\\
\noalign{\smallskip}
$1.20$ -- $1.40$
 &    243
 &    307
 &    275
 &    128
 &     85
 &     11
 &  
\\
 & (     7,    15,   221)
 & (     6,    13,   288)
 & (     2,     8,   265)
 & (     1,     3,   124)
 & (     0,     3,    82)
 & (     0,     0,    11)
 &  
\\
 &     7.67$_{-0.39}^{+0.25}$
 &     7.77$_{-0.08}^{+0.09}$
 &     7.79$_{-0.06}^{+0.13}$
 &     7.94$_{-0.03}^{+0.10}$
 &     8.08$_{-0.09}^{+0.12}$
 &     8.21$_{-0.11}^{+0.11}$
 &  
\\
\noalign{\smallskip}
$1.40$ -- $1.60$
 &  
 &  
 &    179
 &    206
 &    154
 &     40
 &  
\\
 &  
 &  
 & (     4,    11,   164)
 & (     1,     4,   201)
 & (     0,     4,   150)
 & (    21,    19,     0)
 &  
\\
 &  
 &  
 &     7.82$_{-0.07}^{+0.10}$
 &     8.10$_{-0.18}^{+0.21}$
 &     8.15$_{-0.10}^{+0.10}$
 &     8.32$_{-0.10}^{+0.09}$
 &  
\\
\noalign{\smallskip}
$1.60$ -- $1.80$
 &  
 &  
 &  
 &    158
 &    108
 &    114
 &  
\\
 &  
 &  
 &  
 & (     0,     3,   155)
 & (     1,     3,   104)
 & (    17,    30,    67)
 &  
\\
 &  
 &  
 &  
 &     8.24$_{-0.13}^{+0.18}$
 &     8.30$_{-0.07}^{+0.09}$
 &     8.43$_{-0.07}^{+0.13}$
 &  
\\
\noalign{\smallskip}
$1.80$ -- $2.00$
 &  
 &  
 &  
 &  
 &  
 &    261
 &  
\\
 &  
 &  
 &  
 &  
 &  
 & (     9,     9,   243)
 &  
\\
 &  
 &  
 &  
 &  
 &  
 &     8.49$_{-0.07}^{+0.08}$
 &  
\\
\noalign{\smallskip}
$2.00$ -- $2.25$
 &  
 &  
 &  
 &  
 &  
 &  
 &  
\\
 &  
 &  
 &  
 &  
 &  
 &  
 &  
\\
 &  
 &  
 &  
 &  
 &  
 &  
 &  
\\
\noalign{\smallskip}
$2.25$ -- $2.50$
 &  
 &  
 &  
 &  
 &  
 &  
 &  
\\
 &  
 &  
 &  
 &  
 &  
 &  
 &  
\\
 &  
 &  
 &  
 &  
 &  
 &  
 &  
\\
\noalign{\smallskip}
$2.50$ -- $3.00$
 &  
 &  
 &  
 &  
 &  
 &  
 &  
\\
 &  
 &  
 &  
 &  
 &  
 &  
 &  
\\
 &  
 &  
 &  
 &  
 &  
 &  
 &  
\\
\noalign{\smallskip}
\noalign{\smallskip}\hline  
\end{tabular}

%% file: tab_stat_Mdust_2.txt
\begin{tabular} {cccccccc}
\hline \noalign{\smallskip} 
\multicolumn{8}{c}{$  0.60\leq z<  1.00$}\\
\noalign{\smallskip}\hline\noalign{\smallskip}
$\rm \log SFR$ & \multicolumn{7}{c}{$\rm \log M_{star}$}\\
\noalign{\smallskip}\noalign{\smallskip}
 & 9.75 -- 10.00
 & 10.00 -- 10.25
 & 10.25 -- 10.50
 & 10.50 -- 10.75
 & 10.75 -- 11.00
 & 11.00 -- 11.50
 & 11.50 -- 12.00
 \\ 
\noalign{\smallskip}\hline\noalign{\smallskip}
\noalign{\smallskip}
$-0.75$ -- $-0.25$
 &  
 &  
 &  
 &  
 &  
 &  
 &  
\\
 &  
 &  
 &  
 &  
 &  
 &  
 &  
\\
 &  
 &  
 &  
 &  
 &  
 &  
 &  
\\
\noalign{\smallskip}
$-0.25$ -- $0.25$
 &  
 &  
 &  
 &  
 &  
 &  
 &  
\\
 &  
 &  
 &  
 &  
 &  
 &  
 &  
\\
 &  
 &  
 &  
 &  
 &  
 &  
 &  
\\
\noalign{\smallskip}
$0.25$ -- $0.50$
 &  
 &  
 &  
 &  
 &  
 &  
 &  
\\
 &  
 &  
 &  
 &  
 &  
 &  
 &  
\\
 &  
 &  
 &  
 &  
 &  
 &  
 &  
\\
\noalign{\smallskip}
$0.50$ -- $0.75$
 &  
 &  
 &  
 &  
 &  
 &  
 &  
\\
 &  
 &  
 &  
 &  
 &  
 &  
 &  
\\
 &  
 &  
 &  
 &  
 &  
 &  
 &  
\\
\noalign{\smallskip}
$0.75$ -- $1.00$
 &  
 &  
 &  
 &  
 &  
 &  
 &  
\\
 &  
 &  
 &  
 &  
 &  
 &  
 &  
\\
 &  
 &  
 &  
 &  
 &  
 &  
 &  
\\
\noalign{\smallskip}
$1.00$ -- $1.20$
 &  
 &    643
 &  
 &     39
 &  
 &     99
 &  
\\
 &  
 & (     5,    15,   623)
 &  
 & (     1,     6,    32)
 &  
 & (     1,     9,    89)
 &  
\\
 &  
 &     7.53$_{-0.19}^{+0.44}$
 &  
 &     7.79$_{-0.38}^{+0.60}$
 &  
 &     7.36$_{-0.22}^{+0.57}$
 &  
\\
\noalign{\smallskip}
$1.20$ -- $1.40$
 &    557
 &    894
 &    385
 &    170
 &     34
 &     12
 &  
\\
 & (    10,    16,   531)
 & (     9,    27,   858)
 & (     5,     9,   371)
 & (     2,     6,   162)
 & (     2,     8,    24)
 & (     0,     1,    11)
 &  
\\
 &     7.64$_{-0.33}^{+0.48}$
 &     7.50$_{-0.20}^{+0.33}$
 &     7.56$_{-0.14}^{+0.16}$
 &     7.69$_{-0.10}^{+0.04}$
 &     7.83$_{-0.15}^{+0.38}$
 &     7.93$_{-0.12}^{+0.29}$
 &  
\\
\noalign{\smallskip}
$1.40$ -- $1.60$
 &  
 &    513
 &    626
 &    351
 &    183
 &     30
 &  
\\
 &  
 & (     7,    15,   491)
 & (     8,    15,   603)
 & (     5,     2,   344)
 & (     2,     6,   175)
 & (     2,     0,    28)
 &  
\\
 &  
 &     7.67$_{-0.15}^{+0.18}$
 &     7.67$_{-0.11}^{+0.11}$
 &     7.80$_{-0.10}^{+0.19}$
 &     7.96$_{-0.10}^{+0.24}$
 &     8.03$_{-0.13}^{+0.10}$
 &  
\\
\noalign{\smallskip}
$1.60$ -- $1.80$
 &  
 &    146
 &    594
 &    462
 &    271
 &     34
 &     42
\\
 &  
 & (     1,     4,   141)
 & (     7,    25,   562)
 & (     6,     7,   449)
 & (     3,     5,   263)
 & (     0,     0,    34)
 & (     2,     4,    36)
\\
 &  
 &     7.97$_{-0.20}^{+0.34}$
 &     8.01$_{-0.16}^{+0.24}$
 &     8.07$_{-0.06}^{+0.09}$
 &     8.25$_{-0.11}^{+0.10}$
 &     8.26$_{-0.10}^{+0.10}$
 &     8.47$_{-0.26}^{+0.62}$
\\
\noalign{\smallskip}
$1.80$ -- $2.00$
 &  
 &  
 &    383
 &    387
 &    328
 &     18
 &     69
\\
 &  
 &  
 & (     5,    10,   368)
 & (     4,    16,   367)
 & (     2,     3,   323)
 & (     0,     0,    18)
 & (     5,     4,    60)
\\
 &  
 &  
 &     8.26$_{-0.20}^{+0.29}$
 &     8.26$_{-0.13}^{+0.14}$
 &     8.32$_{-0.08}^{+0.13}$
 &     8.47$_{-0.03}^{+0.07}$
 &     8.74$_{-0.31}^{+0.46}$
\\
\noalign{\smallskip}
$2.00$ -- $2.25$
 &  
 &  
 &    117
 &    167
 &    310
 &     16
 &     40
\\
 &  
 &  
 & (     3,     7,   107)
 & (     2,     5,   160)
 & (     2,     4,   304)
 & (     0,     0,    16)
 & (     1,     0,    39)
\\
 &  
 &  
 &     8.72$_{-0.18}^{+0.13}$
 &     8.13$_{-0.04}^{+0.13}$
 &     8.39$_{-0.09}^{+0.10}$
 &     8.57$_{-0.05}^{+0.08}$
 &     8.70$_{-0.13}^{+0.21}$
\\
\noalign{\smallskip}
$2.25$ -- $2.50$
 &  
 &  
 &  
 &  
 &  
 &     12
 &  
\\
 &  
 &  
 &  
 &  
 &  
 & (     0,     0,    12)
 &  
\\
 &  
 &  
 &  
 &  
 &  
 &     8.75$_{-0.09}^{+0.15}$
 &  
\\
\noalign{\smallskip}
$2.50$ -- $3.00$
 &  
 &  
 &  
 &  
 &  
 &  
 &  
\\
 &  
 &  
 &  
 &  
 &  
 &  
 &  
\\
 &  
 &  
 &  
 &  
 &  
 &  
 &  
\\
\noalign{\smallskip}
\noalign{\smallskip}\hline  
\end{tabular}

%% file: tab_stat_Mdust_3.txt
\begin{tabular} {cccccccc}
\hline \noalign{\smallskip} 
\multicolumn{8}{c}{$  1.00\leq z<  1.50$}\\
\noalign{\smallskip}\hline\noalign{\smallskip}
$\rm \log SFR$ & \multicolumn{7}{c}{$\rm \log M_{star}$}\\
\noalign{\smallskip}\noalign{\smallskip}
 & 9.75 -- 10.00
 & 10.00 -- 10.25
 & 10.25 -- 10.50
 & 10.50 -- 10.75
 & 10.75 -- 11.00
 & 11.00 -- 11.50
 & 11.50 -- 12.00
 \\ 
\noalign{\smallskip}\hline\noalign{\smallskip}
\noalign{\smallskip}
$-0.75$ -- $-0.25$
 &  
 &  
 &  
 &  
 &  
 &  
 &  
\\
 &  
 &  
 &  
 &  
 &  
 &  
 &  
\\
 &  
 &  
 &  
 &  
 &  
 &  
 &  
\\
\noalign{\smallskip}
$-0.25$ -- $0.25$
 &  
 &  
 &  
 &  
 &  
 &  
 &  
\\
 &  
 &  
 &  
 &  
 &  
 &  
 &  
\\
 &  
 &  
 &  
 &  
 &  
 &  
 &  
\\
\noalign{\smallskip}
$0.25$ -- $0.50$
 &  
 &  
 &  
 &  
 &  
 &  
 &  
\\
 &  
 &  
 &  
 &  
 &  
 &  
 &  
\\
 &  
 &  
 &  
 &  
 &  
 &  
 &  
\\
\noalign{\smallskip}
$0.50$ -- $0.75$
 &  
 &  
 &  
 &  
 &  
 &  
 &  
\\
 &  
 &  
 &  
 &  
 &  
 &  
 &  
\\
 &  
 &  
 &  
 &  
 &  
 &  
 &  
\\
\noalign{\smallskip}
$0.75$ -- $1.00$
 &  
 &  
 &  
 &  
 &  
 &  
 &  
\\
 &  
 &  
 &  
 &  
 &  
 &  
 &  
\\
 &  
 &  
 &  
 &  
 &  
 &  
 &  
\\
\noalign{\smallskip}
$1.00$ -- $1.20$
 &  
 &  
 &  
 &  
 &  
 &  
 &  
\\
 &  
 &  
 &  
 &  
 &  
 &  
 &  
\\
 &  
 &  
 &  
 &  
 &  
 &  
 &  
\\
\noalign{\smallskip}
$1.20$ -- $1.40$
 &  
 &  
 &  
 &    637
 &  
 &  
 &  
\\
 &  
 &  
 &  
 & (     9,    26,   602)
 &  
 &  
 &  
\\
 &  
 &  
 &  
 &     7.51$_{-0.18}^{+0.69}$
 &  
 &  
 &  
\\
\noalign{\smallskip}
$1.40$ -- $1.60$
 &  
 &     14
 &    253
 &    772
 &    669
 &    439
 &  
\\
 &  
 & (     1,     0,    13)
 & (     5,     5,   243)
 & (     8,    22,   742)
 & (     8,    16,   645)
 & (     7,    14,   418)
 &  
\\
 &  
 &     7.89$_{-0.40}^{+0.52}$
 &     7.97$_{-0.28}^{+0.27}$
 &     8.01$_{-0.24}^{+0.16}$
 &     8.16$_{-0.38}^{+0.39}$
 &     7.85$_{-0.14}^{+0.37}$
 &  
\\
\noalign{\smallskip}
$1.60$ -- $1.80$
 &  
 &     17
 &     91
 &    497
 &    531
 &    501
 &     16
\\
 &  
 & (     6,    11,     0)
 & (     0,     2,    89)
 & (     5,     8,   484)
 & (     7,    14,   510)
 & (     5,     6,   490)
 & (     0,     1,    15)
\\
 &  
 &     7.64$_{-0.07}^{+0.42}$
 &     7.93$_{-0.17}^{+0.20}$
 &     7.85$_{-0.04}^{+0.17}$
 &     8.03$_{-0.10}^{+0.17}$
 &     8.23$_{-0.09}^{+0.21}$
 &     8.31$_{-0.14}^{+0.20}$
\\
\noalign{\smallskip}
$1.80$ -- $2.00$
 &  
 &  
 &  
 &    196
 &    279
 &    376
 &  
\\
 &  
 &  
 &  
 & (     4,     1,   191)
 & (     1,     6,   272)
 & (     6,     5,   365)
 &  
\\
 &  
 &  
 &  
 &     8.02$_{-0.04}^{+0.20}$
 &     8.21$_{-0.10}^{+0.13}$
 &     8.44$_{-0.07}^{+0.15}$
 &  
\\
\noalign{\smallskip}
$2.00$ -- $2.25$
 &  
 &  
 &     12
 &     36
 &  
 &    114
 &     17
\\
 &  
 &  
 & (     5,     7,     0)
 & (     0,     0,    36)
 &  
 & (     1,     6,   107)
 & (     0,     1,    16)
\\
 &  
 &  
 &     8.34$_{-0.22}^{+0.23}$
 &     8.17$_{-0.06}^{+0.18}$
 &  
 &     8.53$_{-0.09}^{+0.17}$
 &     8.82$_{-0.13}^{+0.18}$
\\
\noalign{\smallskip}
$2.25$ -- $2.50$
 &  
 &  
 &    210
 &    105
 &  
 &     16
 &     10
\\
 &  
 &  
 & (     5,     8,   197)
 & (     3,     6,    96)
 &  
 & (     0,     0,    16)
 & (     4,     6,     0)
\\
 &  
 &  
 &     8.61$_{-0.40}^{+0.36}$
 &     8.39$_{-0.06}^{+0.26}$
 &  
 &     8.76$_{-0.13}^{+0.11}$
 &     8.89$_{-0.21}^{+0.21}$
\\
\noalign{\smallskip}
$2.50$ -- $3.00$
 &  
 &  
 &  
 &  
 &  
 &     19
 &  
\\
 &  
 &  
 &  
 &  
 &  
 & (     0,     0,    19)
 &  
\\
 &  
 &  
 &  
 &  
 &  
 &     9.11$_{-0.17}^{+0.10}$
 &  
\\
\noalign{\smallskip}
\noalign{\smallskip}\hline  
\end{tabular}

%% file: tab_stat_Mdust_4.txt
\begin{tabular} {cccccccc}
\hline \noalign{\smallskip} 
\multicolumn{8}{c}{$  1.50\leq z<  2.50$}\\
\noalign{\smallskip}\hline\noalign{\smallskip}
$\rm \log SFR$ & \multicolumn{7}{c}{$\rm \log M_{star}$}\\
\noalign{\smallskip}\noalign{\smallskip}
 & 9.75 -- 10.00
 & 10.00 -- 10.25
 & 10.25 -- 10.50
 & 10.50 -- 10.75
 & 10.75 -- 11.00
 & 11.00 -- 11.50
 & 11.50 -- 12.00
 \\ 
\noalign{\smallskip}\hline\noalign{\smallskip}
\noalign{\smallskip}
$-0.75$ -- $-0.25$
 &  
 &  
 &  
 &  
 &  
 &  
 &  
\\
 &  
 &  
 &  
 &  
 &  
 &  
 &  
\\
 &  
 &  
 &  
 &  
 &  
 &  
 &  
\\
\noalign{\smallskip}
$-0.25$ -- $0.25$
 &  
 &  
 &  
 &  
 &  
 &  
 &  
\\
 &  
 &  
 &  
 &  
 &  
 &  
 &  
\\
 &  
 &  
 &  
 &  
 &  
 &  
 &  
\\
\noalign{\smallskip}
$0.25$ -- $0.50$
 &  
 &  
 &  
 &  
 &  
 &  
 &  
\\
 &  
 &  
 &  
 &  
 &  
 &  
 &  
\\
 &  
 &  
 &  
 &  
 &  
 &  
 &  
\\
\noalign{\smallskip}
$0.50$ -- $0.75$
 &  
 &  
 &  
 &  
 &  
 &  
 &  
\\
 &  
 &  
 &  
 &  
 &  
 &  
 &  
\\
 &  
 &  
 &  
 &  
 &  
 &  
 &  
\\
\noalign{\smallskip}
$0.75$ -- $1.00$
 &  
 &  
 &  
 &  
 &  
 &  
 &  
\\
 &  
 &  
 &  
 &  
 &  
 &  
 &  
\\
 &  
 &  
 &  
 &  
 &  
 &  
 &  
\\
\noalign{\smallskip}
$1.00$ -- $1.20$
 &  
 &  
 &  
 &  
 &  
 &  
 &  
\\
 &  
 &  
 &  
 &  
 &  
 &  
 &  
\\
 &  
 &  
 &  
 &  
 &  
 &  
 &  
\\
\noalign{\smallskip}
$1.20$ -- $1.40$
 &  
 &  
 &  
 &  
 &  
 &  
 &  
\\
 &  
 &  
 &  
 &  
 &  
 &  
 &  
\\
 &  
 &  
 &  
 &  
 &  
 &  
 &  
\\
\noalign{\smallskip}
$1.40$ -- $1.60$
 &  
 &  
 &  
 &  
 &  
 &  
 &  
\\
 &  
 &  
 &  
 &  
 &  
 &  
 &  
\\
 &  
 &  
 &  
 &  
 &  
 &  
 &  
\\
\noalign{\smallskip}
$1.60$ -- $1.80$
 &  
 &  
 &  
 &  
 &  
 &  
 &  
\\
 &  
 &  
 &  
 &  
 &  
 &  
 &  
\\
 &  
 &  
 &  
 &  
 &  
 &  
 &  
\\
\noalign{\smallskip}
$1.80$ -- $2.00$
 &  
 &  
 &  
 &    120
 &     59
 &     33
 &  
\\
 &  
 &  
 &  
 & (     9,    19,    92)
 & (     1,    11,    47)
 & (     2,     3,    28)
 &  
\\
 &  
 &  
 &  
 &     7.85$_{-0.10}^{+0.26}$
 &     7.94$_{-0.14}^{+0.42}$
 &     8.10$_{-0.07}^{+0.26}$
 &  
\\
\noalign{\smallskip}
$2.00$ -- $2.25$
 &  
 &  
 &    434
 &    615
 &    352
 &    234
 &     20
\\
 &  
 &  
 & (    13,    13,   408)
 & (    12,    23,   580)
 & (     9,    18,   325)
 & (     3,    10,   221)
 & (     0,     1,    19)
\\
 &  
 &  
 &     7.92$_{-0.08}^{+0.42}$
 &     8.07$_{-0.11}^{+0.13}$
 &     8.14$_{-0.07}^{+0.20}$
 &     8.28$_{-0.01}^{+0.12}$
 &     8.46$_{-0.05}^{+0.29}$
\\
\noalign{\smallskip}
$2.25$ -- $2.50$
 &  
 &  
 &    224
 &    617
 &    708
 &    479
 &     29
\\
 &  
 &  
 & (     9,     7,   208)
 & (     5,    25,   587)
 & (     6,    26,   676)
 & (     3,    12,   464)
 & (     1,     1,    27)
\\
 &  
 &  
 &     8.32$_{-0.05}^{+0.27}$
 &     8.34$_{-0.02}^{+0.15}$
 &     8.37$_{-0.02}^{+0.14}$
 &     8.60$_{-0.04}^{+0.18}$
 &     8.78$_{-0.16}^{+0.20}$
\\
\noalign{\smallskip}
$2.50$ -- $3.00$
 &  
 &  
 &  
 &    208
 &    406
 &    544
 &     37
\\
 &  
 &  
 &  
 & (     6,     4,   198)
 & (     6,    10,   390)
 & (     9,    21,   514)
 & (     0,     1,    36)
\\
 &  
 &  
 &  
 &     8.75$_{-0.04}^{+0.21}$
 &     8.71$_{-0.04}^{+0.16}$
 &     8.94$_{-0.03}^{+0.14}$
 &     9.18$_{-0.12}^{+0.14}$
\\
\noalign{\smallskip}
\noalign{\smallskip}\hline  
\end{tabular}